\documentclass[11pt,letterpaper]{article}
\usepackage{jcappub}
\usepackage{caption}
\usepackage{floatrow}
\usepackage{feynmp}

\usepackage{bm}

\global\def\mypath#1#2#3{
 \fmfiset{p#1}{vpath#1(__#2,__#3)}
 \fmfi{wiggly}{subpath (0,length(p#1)/2) of p#1}
 \fmfi{plain}{subpath (length(p#1)/2,length(p#1)) of p#1}
 }

\DeclareNewFloatType{myfig}{placement=H,name=}
\DeclareCaptionLabelFormat{myformat1}{(#2)}
\newenvironment{myfigure}
  {\setlength\floatsep{0pt}\setlength\textfloatsep{0pt}\setlength\intextsep{5pt}\setcounter{myfig}{\value{equation}}
  \numberwithin{myfig}{section}
  \addtocounter{equation}{1}
    \begin{myfig}}
  {\end{myfig}}

\def\be{\begin{eqnarray}}
\def\ee{\end{eqnarray}}

\def\f{\mathrm{\textsl{f}}\,}
\def\k{\bm{k}}
\def\o{\bm{w}}
\def\d{\delta_D}
\def\c{\mathrm{c}}
\def\la{\langle}
\def\ra{\rangle}
\def\dD{\mathcal{D}}
\def\R{\mathcal{R}}
\def\vf{\varphi}
\def\l{\lambda}
\def\t{\tau}
\renewcommand{\L}[2]{L_{#1 #2}}
\newcommand{\RR}[2]{\overrightarrow{R_{#1 #2}}}
\newcommand{\RRl}[2]{\overleftarrow{R_{#1 #2}}}
\newcommand{\inv}[1]{\left[#1^{-1}\right]}

\newcommand{\OK}[1]{\mathcal{O}\left((Kf^2)^{#1}\right)}

\title{ The Helmholtz Hierarchy: \\Phase Space Statistics of Cold
Dark Matter}
\author{Svetlin V. Tassev}
\affiliation{Center for Astrophysics, Harvard University\\
Cambridge, MA 02138, USA}
\abstract{We present a new formalism to study large-scale structure in the universe. The result is a hierarchy (which we call the ``Helmholtz Hierarchy'') of equations describing the phase space statistics of cold dark matter (CDM). The hierarchy features a physical ordering parameter which interpolates between the Zel'dovich approximation and fully-fledged gravitational interactions. The results incorporate the effects of stream crossing. We show that the Helmholtz hierarchy is self-consistent and obeys causality to all orders. We present 
 an interpretation of the hierarchy in terms of effective particle trajectories.} 

\notoc
\begin{document}
\maketitle

\section{Introduction}\label{sec:introd}

Understanding the evolution of large-scale structure in the universe is an important ingredient of present-day cosmology. Neglecting the complications arising from the presence of baryons, the evolution of large-scale structure is governed entirely by the dynamics of Cold Dark Matter (CDM). Even though the Vlasov and Poisson equations specify completely the dynamics of CDM at scales much smaller than the horizon, structure formation is still not well-understood since the dynamics is both nonlinear and non-local due to the long range of the gravitational interaction.

 CDM evolution in the weakly nonlinear and nonlinear regimes so far has been studied efficiently only through numerical simulations. Accurate predictions of the CDM correlation functions in the mildly nonlinear regime ($k\sim0.1\,$Mpc$^{-1}$ at $z\sim0$) will be important for comparison with ongoing and future experiments, such as BOSS\footnote{http://www.sdss3.org/surveys/boss.php} and WFIRST\footnote{http://wfirst.gsfc.nasa.gov}, targeting the baryon acoustic oscillations (BAO) in the matter power spectrum in order to put constraints on dark energy models. The numerical simulations conducted so far trying to assess the nonlinear effects on the predictions from linear perturbation theory, have suffered from one or more of the following: 1) low mass resolution (thus requiring phenomenological halo bias to be used for modeling the galaxy distribution, e.g. \cite{Cole:1998vh}); 2) sampling variance (e.g. \cite{Angulo:2007fw}); 3) studies with small simulation boxes miss effects from the nonlinear evolution of longer wavelength modes \cite{Crocce:2005xz}; 4) in numerical simulations one usually fixes all cosmological parameters to infinite precision, and then varies only the dark energy equation of state, which is not adequate for getting a complete handle on the uncertainties (e.g. \cite{Angulo:2007fw}). 

Therefore, apart from the numerical work, numerous analytical methods have been employed to study the CDM power spectrum. However, all analytical expansion schemes used so far fail to achieve 1\% accuracy for the density power spectrum even in the weakly nonlinear regime (see below). This is due to the fact that the convergence properties of these expansion schemes rely on the existence of a small parameter, and such a parameter simply does not exist in the nonlinear regime. Thus, even for the simplified dynamics in the Zel'dovich approximation (ZA) \cite{zeldovich} all consistent Eulerian analytical expansion schemes fail to recover the density power spectrum when the fractional overdensity becomes close to one \cite{Valageas:2007ge}.

Moreover, most expansion schemes work in the single-stream approximation. Thus, the results of these methods are applicable before shell-crossing, i.e. long before virialization can occur. This is done by considering only the first two moments of the Vlasov equation which reduce to the usual continuity and Euler equations. Since the higher moments of the one-particle distribution function are artificially discarded, one closes the system by introducing an equation of state, or equivalently a sound speed, which is set to zero for the CDM (along with any anisotropic stress). 

All analytical methods based on the single-stream approximation effectively neglect high-$k$ modes, some of which collapse (and therefore develop shell-crossings) as soon as one starts evolving the full Vlasov-Poisson (VP) system. It was recently showed \cite{Baumann:2010tm} that virialized structures decouple completely from large scale modes, just as the collision of two galaxies is not affected on galactic scales if all stars were substituted by binaries (i.e. high-$k$ virialized objects). Yet, after averaging over the nonlinear structures, the dynamics of the low-$k$ modes is slightly modified due to the presence of non-virialized structures by introducing a non-vanishing effective sound speed and viscosity \cite{Baumann:2010tm}. In the language of statistical mechanics, there is an effective field theory which governs the behavior of the low-$k$ modes, when one integrates out the high-$k$ degrees of freedom. 

Recently various analytical tricks have been put forward to remedy the convergence properties in the weakly nonlinear regime of the standard Eulerian perturbation theory \cite{jain}. These include renormalized perturbation theory \cite{Crocce:2005xy}, the path-integral approach \cite{Valageas:2006bi}, and the renormalization group flow \cite{Matarrese:2007aj}. A critique to these methods, which rely on the single-stream approximation, was presented by \cite{Afshordi:2006ch} (see \cite{Valageas:2010rx} as well), showing that at $z=0$ for $k\gtrsim 0.1\,$Mpc$^{-1}$ shell-crossing may affect the power spectrum at a percent level, which is comparable to the projected observational errors of future BAO experiments. A similar result was obtained by \cite{Carlson:2009it}, who demonstrate that all of the existing analytical techniques deviate at 1\% or more from the correct power spectrum derived from simulations for $k\gtrsim 0.1\,$Mpc$^{-1}$ at $z=0$. Moreover, \cite{Carlson:2009it} find that those methods systematically fail to reproduce the density-velocity cross-correlation for the same scales. 

Approximating the CDM as a pressureless perfect fluid implies that one discards any short-scale velocity dispersion, which is generated through the nonlinear evolution, as shown in e.g. \cite{McDonald:2009hs}. By obtaining the stress tensor from numerical simulations, it was shown in \cite{Pueblas:2008uv} that neglecting the velocity dispersion can lead to 1\% effect on the power spectrum for $k\gtrsim 0.2\,$Mpc$^{-1}$ at $z=0$. A systematic analy\-sis of different expansion schemes using the ZA was performed by \cite{Valageas:2007ge} who shows that for $k\gtrsim 0.1\,$Mpc$^{-1}$ there is an agreement (to better than 1\%) with the exact nonlinear power only for expansions which are not self-consistent, or which add an ansatz for the decay at high $k$ of the \textit{density} response function, which may not be adequate for the gravitational dynamics. Therefore, if one wants to achieve 1\% accuracy at the scales relevant for BAO measurements, one needs to go beyond the pressureless perfect fluid approximation, and consider extending the analysis to include the rich structure of the CDM one-particle distribution function in phase space.

Lagrangian Perturbation Theory (LPT) \cite{matsubara}, of which the ZA is a special case (it is the lowest order in LPT), tries to do just that: include the phase-space information in the calculation of the statistical properties of CDM. However, LPT assumes that the velocity field in each CDM stream is irrotational in Eulerian space (see e.g. \cite{Catelan:1994ze}), which is violated after shell crossing. Moreover, it still uses the overdensity as an expansion parameter (albeit in Lagrangian space), which becomes $\mathcal{O}(1)$ at the scales relevant to the BAO. Furthermore, the ``standard'' LPT still assumes the single-stream approximation at intermediate steps, when deriving the second and higher orders in LPT (e.g. \cite{Catelan:1994ze}). Thus, alternatives must be devised.

Another machinery to study CDM statistics in phase-space is the BBGKY hierarchy \cite{peebles} which couples the $n$-point phase-space correlation functions in an infinite hierarchy of partial differential equations (PDEs). However, the BBGKY hierarchy suffers from a severe closure problem as there is no manifest small physical ordering parameter controlling the hierarchy in the mildly non-linear and non-linear regimes (see \cite{peebles} for further discussion). 

Another line of research was followed by Valageas \cite{Valageas:2003gm} who used the steepest-descent method applied to a large-$N$ expansion to obtain equations of motion for the phase-space statistics of the CDM. He used an expansion in the linear power spectrum $P_L$, which recovers the results of standard perturbation theory, and so does not give insight into the effects of stream crossing on the power spectrum.\footnote{We postpone a further discussion of his results to Sections \ref{sec:zeldovich_IC} and \ref{sec:summary_disc_6}.}

The aim of this paper is to start from first principles and obtain equations of motion governing the evolution of the phase-space statistics of CDM. This will result in a hierarchy, which we call the Helmholtz hierarchy (HH), which is self-consistent to all orders, obeys causality, and has a physical ordering parameter. 

One can recast the CDM dynamics into a problem in statistical mechanics, since the initial conditions for structure formation are such that the density field (along with the particle positions and velocities) is a stochastic random field. Thus, each realization of that field can be treated as a microstate and the resulting canonical ensemble can be described by a well-defined Gibbs partition function \cite{Valageas:2003gm}. From the partition function one can obtain the Gibbs and Helmholtz free energies, which generate the extended BBGKY hierarchy and the Helmholtz hierarchy (both defined below), respectively. 

Since structure formation is an inherently out-of-equilibrium process, no simple relation (analogous to the fluctuation-dissipation theorem) holds between the correlation and response functions of the CDM. Instead, the $n$-point correlation functions and the $n$-point response functions of $m$-th order are coupled through what we call the ``extended'' BBGKY hierarchy (see Section \ref{sec:BBGKY}). However, as in the case with the standard BBGKY hierarchy, the extended BBGKY hierarchy has no manifest small physical ordering parameter in the mildly non-linear and non-linear regimes.

The Helmholtz hierarchy, similar to the BBGKY hierarchy, describes the phase-space statistics of CDM.\footnote{The HH governs the evolution of the functional derivatives of the Helmholtz free energy, which is also called the 1 Particle Irreducible (1PI) effective action, $\Gamma$, defined later for the CDM. Those quantities are related to the phase-space correlation and response functions of CDM through a Legendre transformation.}
However, we show that the Helmholtz hierarchy is regulated by a physical ordering parameter, which is schematically given by the fractional difference between the acceleration of test particles as given by a Zel'dovich-type approximation (i.e. the acceleration is assumed parallel to the velocity at an intermediate moment in time), and their corresponding true acceleration due to gravity. Therefore, it effectively interpolates between Zel'dovich dynamics and fully-fledged gravitational dynamics.  

We will see that under a sharp truncation of the HH all $n$-point correlation functions of CDM are generated, in stark contrast to the (extended) BBGKY hierarchy.  Combining this result with the presence of a physical ordering parameter, we will show that the HH ameliorates the closure problem of the (extended) BBGKY hierarchy.


By constructing the above hierarchy we showed that the effects of stream crossing are not non-perturbative as suggested by some authors (e.g. \cite{Valageas:2003gm}), but are the result of carefully treating the initial conditions in phase-space. We show that although the initial overdensities follow Gaussian statistics, the initial one-particle distribution function in phase-space has highly non-Gaussian features, preserving which is crucial for the consistency of the method, and for capturing the correct physics as well. 

One can construct closed-form solutions for the Helmholtz hierarchy at each order, and show that they are closely related to a Born-type solution to the Vlasov-Poisson equation. By doing that to the second lowest order and reabsorbing some higher order contributions, we show that the Helmholtz hierarchy (HH) solutions have a natural interpretation. One can think of the solutions in terms of an iterative improvement of the ZA using N-body simulations in the following way:

\textbf{Initial set-up:} One starts with a simulation box with an initially uniform distribution of CDM particles. One then allows those particles to move according to the ZA, i.e. on exactly solvable straight trajectories. The resulting density field from those particles produces a gravitational potential $\Phi_0$, which does not influence the particles, but is stored for the next iteration step.

\textbf{For $\bm{n=1,2, ...}$ do:} A new simulation box is constructed which is filled with exactly the same particles at the same initial positions as in the initial set-up\footnote{This is done so that one works in the same realization of the initial conditions.}. One then solves the standard equations of motion for each particle, but for the gravitational acceleration one uses\footnote{One has a choice of whether to evaluate the acceleration of each particle, $(-\nabla \Phi_{n-1})$, at the position of that particle in the box of the $(n-1)-$th iteration, or at the position of that particle in the box of the $n-$th iteration. Further analysis is necessary to establish which approach yields faster converging results.} $(-\nabla \Phi_{n-1})$. The resulting density field from those particles produces a gravitational potential $\Phi_n$, which does not influence the particles, but is stored for the next iteration step.

At each iteration step, the above scheme produces density fields which gradually interpolate between the ZA and the fully non-linear solution. The gravitational potential resulting from the ZA is much smoother than the result in the fully non-linear regime, thus resulting in much smaller accelerations, $\bm{a}$, especially at small scales. Those small-scale particle accelerations gradually increase with each iteration. Since the usual choice for timesteps in N-body simulations goes like $1/\sqrt{|\bm{a}|}$, one may hope that even if the integrals involved in the Helmholtz solutions are not doable numerically, the above N-body scheme will produce the desired solutions using very few iterations with very few timesteps each\footnote{One needs zero timesteps in the case of the ZA, which is exactly solvable. One can also imagine numerous other improvements to the above iteration scheme, such as using the second order LPT solution for the initial set-up, which has been shown to improve the behavior of full-blown N-body simulations \citep{Crocce:2006ve}. Another possible optimization would be achieved by filtering out the small scales in the initial conditions. Their dynamics cannot be captured by the ZA, and would require a large number of iterations to recover using the above iterative scheme. Those scales are hardly relevant, because small scale non-linear power should have almost no effect on the mildly non-linear scales \cite{peebles}.} to recover the mildly non-linear regime. This can result in a speed-up of N-body simulations targeting that regime. The results can have a wider applicability than simply calculating the matter power spectrum, since other statistics can be extracted from the simulation boxes in the same way one treats the results from fully non-linear N-body simulations.


Our analysis mainly builds upon the work of \cite{Valageas:2003gm} and \cite{Gasenzer:2007}. Thus, most of the technical complications in obtaining the HH arise because the action governing the classical dynamics of CDM in the presence of the stochastic initial conditions is built up of non-trivial integro-differential operators. To aid the reader, we therefore summarize the results at the end of each of the more technical sections.

In Section \ref{sec:prelims} we start with a review of the Zel'dovich approximation, deriving the phase-space response and correlation functions of the CDM in the ZA. We then write down the full VP equation and in Section \ref{sec:BBGKY} we derive the extended BBGKY hierarchy, which includes the response functions of the CDM. Before we can derive the HH, we have to rewrite the Vlasov-Poisson system, given the stochastic initial conditions set by inflation, in the form of a path integral. This is done in Section \ref{sec:VP}. Next we review the Non-Perturbative Renormalization Group (NPRG) flow equations in Section \ref{sec:NPRG}, which we then solve for the CDM to obtain the Helmholtz hierarchy in Section \ref{sec:SOL}. We reintroduce the non-Gaussianities of the initial one-particle distribution function in Section \ref{app:ngf}. And in Section \ref{PIinterp} we show how one can write the HH in terms of an iterative scheme of N-body simulations. We then summarize our results in Section \ref{sec:summary}.

\section{Preliminaries}\label{sec:prelims}
\subsection{The Zel'dovich approximation}
We begin our discussion with an overview of the ZA \cite{zeldovich}. It is the lowest order approximation in Lagrangian perturbation theory and by construction in the Helmholtz Hierarchy as well; and it provides the basis for the expansion which is at the root of the HH. In this section we derive the one-point and two-point correlation functions in the ZA, as well as the linear response function. This will give us a flavor of the phase-space statistics of CDM. Although the ZA breaks down in the nonlinear regime, it captures some of the effects of stream crossing of CDM, and thus it will give us an insight which is qualitatively different from the results in the single-stream approximation. 

In the ZA, the Eulerian comoving coordinates, $\bm{x}$, of a particle are given by
\be
\bm{x}(\bm{q},\eta)=\bm{q}+D(\eta)\bm{s}(\bm{q})\ .
\ee
We denote conformal time with $\eta$; $\bm{q}$ are the Lagrangian coordinates; $\bm{s}$ is a stochastic displacement field, encoding the initial conditions for structure formation; and $D$ is the linear growth factor\footnote{The linear growth factor is given by the growing solution of $d(a\dot D)/d\eta=4\pi G \bar\rho_M D$ with $D(\eta_0)=1$ today (at $\eta_0$), where $G$ is the Newton constant, $\bar \rho_M$ is the average comoving matter density, and $a$ is the cosmological scale factor. }. The above equation and eq.~(\ref{s}) below can be derived in linear theory (e.g. \cite{Valageas:2007ge}). In the ZA they are assumed to hold in the nonlinear regime as well. 

The stochastic displacement field in Fourier space is given by\footnote{We define the Fourier transform as $$\bm{s}(\bm{k})=\int \frac{d^3 q}{(2\pi)^3} e^{-i \bm{q}\cdot\bm{k}} \bm{s}(\bm{q})\ .$$ For brevity we use the same notation for quantities in Fourier and in configuration/phase space. }:
\be\label{s}
\bm{s}(\bm{k})=i\frac{\bm{k}}{k^2}\delta_L(\bm{k})\ ,
\ee
where $\bm{k}$ is the Fourier wavevector corresponding to $\bm{x}$; and $\delta_{L}\equiv\delta(\eta_I)/D(\eta_I)$ is the linear fractional matter density perturbation evaluated today, and $\eta_I$ is some early time in the linear regime. By $\delta$ we denote the fractional matter overdensity, $\delta\equiv(\rho_M-\bar \rho_M)/\bar\rho_M$, where $\bar \rho_M$ is the average comoving matter density.  Throughout this paper we assume that $\delta_{L}$ is drawn from a Gaussian random field. This assumption can be dropped easily, but it will make the analysis much more cumbersome. The linear power spectrum, $P_L(k)$, of $\delta_{L}$ is given by:
\be
\la\delta_{L}(\bm{k})\delta_L(\bm{k}')\ra=\d(\k+\k')P_{L}(k)\ ,
\ee
where $\d(x)$ is the Dirac delta function; angular brackets denote ensemble averaging.

We are now in position to write down the one-particle phase space distribution function. Before we do that, however, we have to decide on a velocity coordinate. The conjugate velocity to $\bm{x}$ is given by ($a$ denotes the cosmological scale factor)
\be\label{vc}
\mathrm{\bf{v}}\equiv a \frac{d\bm{x}}{d\eta}\ ,
\ee
such that $d^3xd^3\mathrm{v}$ is the measure in the one-particle phase space\footnote{We have dropped a constant factor of the CDM (particle mass)$^3$. And in general, by phase space we mean the $(\bm{x},\mathrm{\bf{v}})$ space.}. However, we would like to use a velocity coordinate, $\bm{v}$, such that in the ZA we have an equation of motion given by $d\bm{v}/d\eta=0$ (The reasons for this choice will become apparent when we discuss the full VP equation). The coordinate velocity of the CDM particles in the ZA is $$\frac{d\bm{x}}{d\eta}=\dot D\bm{s}(\bm{q})\ ,$$  where a dot denotes a partial derivative with respect to conformal time.  
Therefore, we work with the rescaled velocity 
\be\label{dxdt}
\bm{v}\equiv (a\,\dot D)^{-1}\mathrm{\bf{v}}\ ,
\ee
which for a test particle in the ZA reduces to $\bm{v}=\bm{s}(\bm{q})$, which is time-independent by construction.

Along with the velocity coordinate, we transform the one-particle distribution itself. So, instead of the invariant one-particle phase space distribution function, $\f(\bm{x},\mathrm{\bf v},\eta)$, we use $f(\bm{x},{\bm v},\eta)$ defined as:
\be\label{f_def}
f(\bm{x},\bm{v},\eta)\equiv\left (a\dot D\right)^{3}\f(\bm{x},a\dot D\bm{ v},\eta)\ .
\ee
Thus, the number of particles in a phase-space element is proportional to $$\f d^3x\,d^3\mathrm{v}=fd^3x\,d^3v\ .$$
In (double) Fourier space, this transformation reads:
\be\label{f_equiv_in_fourier}
f(\k,\o,\eta)&=&\int\frac{d^3x\,d^3v}{(2\pi)^6}e^{-i(\k\cdot\bm{x}+\o\cdot\bm{v})}f(\bm{x},{\bm v},\eta)=\int\frac{d^3x\,d^3\mathrm{v}}{(2\pi)^6}e^{-i(\k\cdot\bm{x}+\mathrm{\bf w}\cdot\mathrm{\bf v})}\f(\bm{x},\mathrm{\bf v},\eta)\nonumber\\
&=&\f\left(\k,\mathrm{\bf w}=\frac{\o}{a\dot D},\eta\right)\ ,
\ee
where $\o\equiv a\dot D\mathrm{\bf w}$ is the wavevector corresponding to $\bm{v}$; while $\mathrm{\bf w}$ corresponds to $\mathrm{\bf v}$.

We can integrate over all streams at a given position $\bm{x}$ to write the exact one-particle phase-space distribution function in the ZA as:
\be\label{f_z}
f(\bm{x},\bm{v},\eta)=\int d^3q\delta_D(\bm{v}-\bm{s}(\bm{q}))\delta_D(\bm{x}-\bm{q}-D(\eta)\bm{s}(\bm{q}))\ .
\ee
Next we Fourier transform $f$ given in (\ref{f_z}) with respect to both $\bm{x}$ and $\bm{v}$: 
\be\label{f_kw}
\nonumber f(\bm{k},\bm{w},\eta)&=&\int \frac{d^3 x d^3 v}{(2\pi)^6} e^{-i (\bm{k}\cdot\bm{x}+\bm{w}\cdot\bm{v})}  f(\bm{x},\bm{v},\eta)  \\
&=&\int\frac{ d^3q}{(2\pi)^6}  e^{-i\bm{k}\cdot\bm{q}}e^{-i\bm{s}(\bm{q})\cdot(D\bm{k}+\bm{w})}\ .
\ee
Thus, the phase-space distribution satisfies the following Vlasov equation:
\be\label{Z_eom}
\dot{f}-\dot D \k\cdot \partial_{\o}f=0\ .
\ee
Now, we are ready to proceed in calculating the correlation and response functions in the ZA.

\subsubsection{Correlation Functions}
Before we proceed to calculate the exact one-point and two-point functions, let us see how $f$ relates to the fractional density perturbation, $\delta$, entering in the single-stream approximation. We have
\be
1+\delta (\bm{x},\eta)=\int d^3\mathrm{ v} \f(\bm{x},\mathrm{\bf v},\eta)=\int d^3v f(\bm{x},\bm{v},\eta)=(2\pi)^3 f(\bm{x},\bm{w}=0,\eta)\ ,
\ee
where we used the partially Fourier transformed $f$. Thus, we find that the CDM density power spectrum is given by
\be
\langle\delta(\bm{k},\eta)\delta(\bm{k}',\eta)\rangle=(2\pi)^6 \langle f(\bm{k},\bm{w}=0,\eta)\,f(\bm{k}',\bm{w}'=0,\eta)\rangle\ .
\ee
In order to calculate the above bracket using the expression for $f$, eq.~(\ref{f_kw}), we use that for any linear combination, $A$, of Gaussian random variables, the following equality for the ensemble average holds:
\be\label{gaussian_expansion}
\langle\exp (A)\rangle=\exp\left(\langle A\rangle_c+\frac{1}{2}\langle A^2\rangle_c\right)\ ,
\ee
where $\langle\rangle_c$ denotes the cumulants.
Thus, we obtain
\be\label{2pfz}
(2\pi)^6 \langle f(\bm{k},\bm{w},\eta)\,f(\bm{k}',\bm{w}',\eta')\rangle=\d(\bm{k}+\bm{k}')\int\frac{d^3 q}{(2\pi)^3}e^{-i\bm{q}\cdot\bm{k}}\times\nonumber\\
\times\exp\left\{-\frac{1}{2}\int d^3\xi \frac{P_L(\xi)}{\xi^4}\mathcal{I}(\bm{q},\bm{\xi},\bm{k},\bm{w},\bm{w}',\eta,\eta')\right\}\, , \ \mathrm{where}\\
\mathcal{I}(\bm{q},\bm{\xi},\bm{k},\bm{w},\bm{w}',\eta,\eta')=\left(\bm{\xi}\cdot\bm{l}\right)^2+\left(\bm{\xi}\cdot\bm{m}\right)^2+
2\cos(\bm{\xi}\cdot\bm{q})\left(\bm{\xi}\cdot\bm{l}\right)\left(\bm{\xi}\cdot\bm{m}\right)\nonumber\\
{\bm l}\equiv {\bm w}+D(\eta){\bm k}\ , \ \  {\bm m}\equiv \bm{w}'-D(\eta')\bm{k}\ .\nonumber
\ee
The above expression is the exact two-point function for CDM in the ZA, which includes the effects of shell-crossing. Setting $\bm{w}=\bm{w}'=0$, the above expression reduces to the expression for the two-point function of the density perturbations (cf. eq.~(105) in \cite{Valageas:2007ge}). 

The above equation can be integrated in $\bm\xi$ to obtain:
\begin{eqnarray}\label{2pfz_simple}
\begin{aligned}
(2\pi)^6 &\langle f(\bm{k},\bm{w},\eta)\,f(\bm{k}',\bm{w}',\eta')\rangle=\\
=\ &\d(\bm{k}+\bm{k}')e^{-\frac{\sigma_v^2}{2}(l^2+m^2)}\times \\
&\times \, \int \frac{d^3 q}{(2\pi)^3} \ \exp\Big[-i{\bm q}\cdot {\bm k}-{\bm l}\cdot{\bm m} \, \big(\chi(q)+\gamma(q)\big)+3 ({\bm l}\cdot {\bm {\hat q}})\,({\bm m}\cdot {\bm {\hat q}}) \gamma(q)\Big]\ ,
\end{aligned}
\end{eqnarray}
where
\be
&\sigma_v^2\equiv \frac{4\pi}{3}\int_0^\infty dx P_L(x)\, , \ 
\chi(q)\equiv \frac{4\pi}{3}\int_0^\infty dx P_L(x)j_0(x q)\, , \ 
\gamma(q)\equiv \frac{4\pi}{3}\int_0^\infty dx P_L(x)j_2(xq)\nonumber\ , \\
\ee
where $j_l$ denote the spherical Bessel functions.

As a consistency check, expanding to linear order in $P_L$ we obtain
\begin{eqnarray}\label{2pfz_simple_expanded}
(2\pi)^6 \langle f(\bm{k},\bm{w},\eta)\,f(\bm{k}',\bm{w}',\eta')\rangle=\ \d(\bm{k}+\bm{k}')P_L(k)\left[-\frac{\left(\bm{l}\cdot\bm{k}\right)\, \left(\bm{m}\cdot\bm{k}\right)}{k^4}\right]\ ,
\end{eqnarray}
which for $\bm{w}=\bm{w}'=0$ and $\eta=\eta'$ reduces to $\langle\delta(\bm{k},\eta)\delta(\bm{k}',\eta)\rangle=\delta^{(3)}(\bm{k}+\bm{k}')P_L(k)D^2$, which is the lowest order expression for the power spectrum in Standard Perturbation Theory (SPT).

Using (\ref{f_kw}) and (\ref{gaussian_expansion}) we can obtain the ensemble averaged one-point function:
\be\label{fbarZ}
\bar{f}=\frac{\d(\bm{k})}{(2\pi)^3}\exp\left[-\frac{\sigma_v^2 w^2}{2}\right]\ ,
\ee
where $\bar f\equiv\langle f\rangle$.
The exponential damping is due to the fact that the large-scale velocity dispersion is given by $\sigma^2_v$. The delta function above encodes the homogeneity and isotropy of the universe.

\subsubsection{Linear response function}
The linear\footnote{By ``linear'' we mean that the response function corresponds to the first functional derivative of $\bar f$ in $\zeta$.} response function, $\delta \bar{f}_a/\delta \zeta_b$, tells us how $\bar{f}_a\equiv\bar{f}(\k_a,\o_a,\eta_a)$ is modified under a forcing $\zeta_b=\zeta(\k_b,\o_b,\eta_b)$ at time $\eta_b$. The forcing can be incorporated in the ensemble averaged Vlasov equation as follows:
\be\label{Z_eom_resp}
\dot{\bar{f}}-\dot D \k\cdot \partial_{\o}\bar{f}=\zeta\ .
\ee
Taking the functional derivative of the above equation, we obtain 
\be
\partial_{\eta_a}\frac{\delta \bar{f}_a}{\delta \zeta_b}-\dot D(\eta_a) \k_a\cdot \partial_{\o_a}\frac{\delta \bar{f}_a}{\delta \zeta_b}=\d(\k_a-\k_b)\d(\o_a-\o_b)\d(\eta_a-\eta_b)\ .
\ee
The solution to this equation is
\be\label{Rabz}
\frac{\delta\bar{f}_a}{\delta \zeta_b}=\theta(\eta_a-\eta_b) \delta_D(\bm{k}_a-\bm{k}_b)\delta_D\bigg(\bm{w}_a-\bm{w}_b+\big(D(\eta_a)-D(\eta_b)\big)\bm{k}_a\bigg)\ ,
\ee 
where $\theta(x)$ is the Heaviside step function. Above we imposed the following conditions \cite{Valageas:2003gm}
\be
\frac{\delta \bar{f}_a}{\delta \zeta_b}\rightarrow \d(\k_a-\k_b)\d(\o_a-\o_b) \ \ \ \hbox{as}\ \ \ \eta_a\rightarrow\eta_b+0^+\nonumber\\
\frac{\delta \bar{f}_a}{\delta \zeta_b}=0\ \ \ \hbox{for}\ \ \ \eta_b>\eta_a\nonumber\\
\frac{\delta \bar{f}(\k_a=0,\o_a=0,\eta_a)}{\delta \zeta_b}=\d(\k_b)\d(\o_b)\ \ \ \hbox{for}\ \ \ \eta_a>\eta_b\ .
\ee
The first two equations are causality conditions: information about the forcing $\zeta$ propagates only forward in time. The third equation comes from mass conservation as $\int d^3xd^3v f=$constant.

As argued by Valageas \cite{Valageas:2007ge}, the high-$k$ damping of the linear \textit{density} response function obtained in Renormalized Perturbation Theory (RPT) \cite{Crocce:2005xz} is due to the erasure of small-scale correlations between the initial and final density field by the advection of high-$k$ structures by the random large-scale flows. Since the high-$k$ decay of the linear \textit{density} response function is due to large-scale bulk motions, it is also present in the ZA \cite{Valageas:2007ge,baradwaj}. However, the exact linear phase-space response function in the ZA, (\ref{Rabz}), does not exhibit such an exponential damping at high $k$. Still, one can recover the exponential damping of the \textit{density} response function by performing an expansion similar to the one done by \cite{matsubara}, but applied directly to the initial conditions of the phase-space distribution function (see also Section \ref{sec:summary_disc_6}).

Having encountered the phase-space statistics of CDM in the simple case of the ZA, we are now ready to turn on the full effects of gravity.

\subsection{The Vlasov-Poisson Equation}\label{sec:zeldovich_IC}

In this paper we work with the full one-particle distribution function $f( \bm{x},\bm{v},\eta)$ for CDM, as defined in (\ref{f_def}). The evolution of $f$ is governed by the coupled Vlasov and Poisson equations, resulting in the VP equation. In later sections, starting from the VP equation, we extract the equations of motion for the phase-space statistics of CDM. Before we do that, however, in this section we write down the VP equation in a convenient form, suitable for the analysis that will follow.

Since we are working with the velocity  defined in (\ref{dxdt}), at early times when the ZA is adequate (${\bm v}\approx\bm{s}(\bm{q})$ for test particles), we have an equation of motion given by $d\bm{v}/d\eta=0$. This will result in the expansion required for the construction of the HH.
The exact equation of motion for a test particle is given by
\be\label{dvdt}
\frac{d\bm{v}}{d\eta}=-\frac{d\ln\left(a\dot D\right)}{d\eta}\,\bm{v}-\frac{1}{\dot D}\partial_{\bm x} \phi\ ,
\ee
where $\phi$ is the Newtonian gravitational potential. The first term above corresponds to the Zel'dovich approximation for the acceleration due to gravity at early times, which is parallel to the velocity. For modes well inside the Hubble horizon, $\phi$ is given by the Poisson equation:
\begin{eqnarray}\label{poisson}
k^2 \phi=-\frac{\delta}{aD}\frac{d}{d\eta}\left(a\dot D\right)\ .
\end{eqnarray}

Using (\ref{vc}), the Vlasov equation for the invariant one-particle distribution function, $\f$, is given by
\be\label{vlasov_std}
\frac{d\f}{d\eta}=\dot \f+\frac{1}{a}\mathrm{\bf v}\cdot\partial_{\bm{x}}\f-a\partial_{\bm{x}} \phi\cdot\partial_{\mathrm{\bf v}}\f=0\ .
\ee
Transforming (\ref{vlasov_std}) according to (\ref{dxdt}) and (\ref{f_def}), and combining with (\ref{poisson}), in Fourier space we obtain
\begin{eqnarray}\label{VP_master}
\dot f&-&\dot D\k\cdot\partial_{\o}f-\\\nonumber
&-&\tilde\varepsilon\frac{d\ln(a\dot D)}{d\eta}\left[(2\pi)^3D^{-1}\int d^3k'\frac{\o\cdot\k'}{k'^2}f(\k',\o=0,\eta)f(\k-\k',\o,\eta)-\o\cdot\partial_{\o}f\right]=0\ .
\end{eqnarray}
This is the Vlasov-Poisson equation. The parameter $\tilde\varepsilon=1$ for the full CDM dynamics, and therefore we drop it altogether through most of our discussion. If one sets $\tilde\varepsilon=0$, one recovers the ZA, (\ref{Z_eom}). Up to a change of variables, the above equation is equivalent to eq.~(7) in \cite{Valageas:2003gm}.


Note that ``Jeans' swindle'' requires that in the above expression we impose \cite{Valageas:2003gm}:
\be\label{jeans}
\frac{\k'}{k'^2}\d(\k')\equiv \frac{\k'}{k'^2+0^+}\d(\k')=0\ .
\ee
Thus, in a homogeneous universe, for which $f(\k,\o,\eta)=(2\pi)^{-3} \d(\k)$, the force term vanishes.

It will prove extremely useful to use deWitt's notation (without taking into account index placement), where subscripts run over all possible labels: i.e. wavevectors, time and possibly fields (several of which appear later in the paper); and repeated subscripts imply summation over discrete labels (e.g. field labels), and integration over continuous labels. 
The VP equation is quadratic in $f_a\equiv f( \k_a,\o_a,\eta_a)$, the quadratic piece coming from the term $\partial_{\bm x} \phi \cdot \partial_{\bm{v}}f$ where $\phi$ is the gravitational potential given by the Poisson equation (\ref{poisson}). 

Thus, the VP equation, (\ref{VP_master}), can be written as:
\be
\L{a}{b} f_b=K_{abc}f_b f_c\ ,
\ee
where $L$ and $K$ are integro-differential operators, defined as follows (no summation below): 
\be
\label{L}
L_{ab}\equiv D_a[\delta_{ab}]\  ,\ \mathrm{where}\\
 D_a\equiv \partial_{\eta_a}-\dot D(\eta_a) \bm{k}_a\cdot \partial_{\bm{w}_a} +\frac{d\ln(a(\eta_a)\dot D(\eta_a))}{d\eta_a}\,\o_a\cdot\partial_{\o_a}\  ,\nonumber\\
\nonumber \delta_{ab}\equiv \delta_D(\bm{k}_a-\bm{k}_b)\d(\bm{w}_a-\bm{w}_b)\d(\eta_a-\eta_b)\ .
\ee
We have defined the operator, $K_{abc}$, to be symmetric in its last two indices (no summation below):
\be
K_{abc}&\equiv& \frac{1}{2} (2\pi)^3\frac{d\ln\left(a(\eta_a)\dot D(\eta_a)\right)}{d\eta_a}D^{-1}(\eta_a)\times\\
&&\times\left[\frac{\k_b\cdot\o_a}{k_b^2}\d(\o_b)\d(\o_a-\o_c)\d(\k_a-\k_b-\k_c)\d(\eta_a-\eta_c)\d(\eta_a-\eta_b)  +\ (b \leftrightarrow c)\right]\ .\nonumber
\label{K}
\ee
Note that the operators above satisfy the following relations: 
\be\label{symm}
\L{a}{b}=\L{-a,}{-b}, \ \mathrm{and}\ K_{abc}=K_{-a,-b,-c}\ ,
\ee 
where
a negative subscript implies that all Fourier wavevectors with that subscript must flip sign. For example, for $\delta_{ab}=\d(\k_a-\k_b)\d(\o_a-\o_b)\d(\eta_a-\eta_b)$ we have $\delta_{a,-b}=\d(\k_a+\k_b)\d(\o_a+\o_b)\d(\eta_a-\eta_b)$. Note that we use a comma in subscripts just as a separator, and not to denote derivatives.

Note that the term in quadratic brackets in eq.~(\ref{VP_master}) vanishes in the ZA, so that eq.~(\ref{VP_master}) reduces to the VP equation in the ZA, eq.~(\ref{Z_eom}). Therefore, we can define an ordering parameter given by  (no summation below):
\be\label{ordering}
\varepsilon_a\equiv\frac{\partial_{\eta_a}\left(\ln\left(a(\eta_a)\dot D(\eta_a)\right)\right)\,\o_a\cdot\partial_{\o_a}f_a-K_{abc}f_bf_c}{\max\{\dot f_a,\dot D(\eta_a) \bm{k}_a\cdot \partial_{\bm{w}_a}f_a\}}\ .
\ee
This is exactly the physical ordering parameter of the HH. Let us use the ZA expression for $f$, (\ref{f_kw}), and see how $\varepsilon$ scales. The numerator vanishes at first order in $\delta$. To second order in $\delta$ in the ZA, we have:
\be
\hbox{numerator of }\varepsilon\sim\partial_{\eta}\ln\left(a\dot D\right) D wks^2\ .
\ee
The denominator to first order in $\delta$ is given by $\sim\dot D \k\cdot \bm{s}$. 

In order to proceed we need an estimate of $w$. The two point function in the ZA can be written as the initial conditions propagated by two response functions until the time of interest (e.g. (\ref{G_Zel})). Since we are interested in quantities such as the density or velocity power spectrum, $G$ is evaluated at $w$ around zero. Then from the response function in the ZA, (\ref{Rabz}), we can see that the largest $w$ we are interested in is\footnote{ If for some reason one is interested in the statistics at very large $w$ (e.g. to calculate statistics involving $\int fv^nd^3v$ with high $n$), the analysis breaks down, since then we would have $w\sim\max\left(Dk ,w_{\hbox{of interest}}\right)\sim w_{\hbox{of interest}}$, and in such a case $\varepsilon$ may not be sufficiently small for an adequate expansion.}  $w\sim D k$. Therefore, we obtain
\be\label{EPS_Z}
\varepsilon\sim \o\cdot\bm{s}\sim D\delta_L\sim\delta \hbox{  in the ZA.}
\ee
In the second equality above we used eq.~(\ref{s}). 

Once we go outside the linear regime, the estimate above breaks down. However, we can still assign a physical significance to $\varepsilon$, starting from the fact that
\be \label{denomEps}
\hbox{denominator of $\varepsilon$ }\sim Kf^2\ .
\ee
 One can check (\ref{denomEps}) both for virialized objects and 
for linear perturbations. For linear perturbations we have $$|Kf^2|\sim (\dot a/a)|\o\cdot\bm{s}|\sim|\dot f|\ ,$$ where we used $w\sim D k$, and $\dot D/D\sim \dot a/a$. While for virialized objects, we have $Kf^2\sim -\dot D \k\cdot\partial_{\o}f$. An easy way to see that is to realize that for a quasistatic halo $\dot f$ vanishes, and the term containing $\o\cdot\partial_{\o}f$ is suppressed, as it is proportional to acceleration in the ZA (i.e. it measures effects of the Hubble flow on the scale of the halo), while $Kf^2$ is proportional to the acceleration of gravity in the halo.\footnote{For the isothermal sphere halo profile, for example, one obtains $$\left|\partial_{\eta}\ln\left(a\dot D\right)\frac{\o\cdot\partial_{\o}f}{Kf^2}\right|\sim \frac{(\hbox{halo length scale})\times \dot a/a}{\hbox{halo velocity dispersion}}\ll 1 .$$}

Combining (\ref{ordering}) with (\ref{denomEps}) we can see that $\varepsilon$ can be schematically represented by the fractional difference between the acceleration of test particles as given by a ZA (as can be seen in the first term in the numerator), and their corresponding true acceleration due to gravity (present in the second term in the numerator). The ordering parameter for the HH effectively interpolates between Zel'dovich-type dynamics and fully-fledged gravitational interactions. Since $K$ is first order in $\varepsilon$, we can see that a truncation to $\mathcal{O}(\varepsilon^{n_\varepsilon})$ requires  a truncation to order $\OK{n_K}$, with $n_K=n_\varepsilon$. Note that $\tilde\varepsilon$ in (\ref{VP_master}) was inserted, so that we can easily keep track of the order of the expansion: A truncation at $\mathcal{O}(\tilde\varepsilon^n)$ is equivalent to a truncation at $\mathcal{O}(\varepsilon^n)$, and vice versa.

It will prove useful to incorporate in the VP equation the initial conditions  for $f$ at time $\eta_I$ (no summation below): 
\be\label{fI}
f_{I,a}\equiv  f(\k_a,\o_a,\eta_I) \d(\eta_a-\eta_I)\ . 
\ee
 As we will see in Section \ref{sec:BBGKY} the resulting modified VP equation will modify the standard  BBGKY hierarchy \cite{peebles} to include the response functions of the CDM dynamics. The modified VP equation reads:
\be\label{vp_stoch}
\L{a}{b} f_b-K_{abc}f_b f_c=f_{I,a}\ .
\ee
This way the initial condition for this modified equation is $f(\eta<\eta_I)=0$. In the case of CDM, $f(\k_a,\o_a,\eta_I)$ is the phase-space distribution obtained in the linear regime after matter-radiation decoupling. Therefore it can be obtained with linear theory from the inflationary initial conditions, and thus, the forcing term, $f_I$, is a stochastic random variable, which acts only at $\eta_I$. 

The first two cumulants of $f_I$ are given by:
\be\label{ICcum}
\mu_a\equiv\langle f_{I,a}\rangle \ \ \hbox{and}\ \ \Delta_{ab}\equiv\langle f_{I,a}f_{I,b}\rangle_\c\ ,
\ee
where angular brackets denote ensemble average over the stochastic initial conditions set by inflation, and we denoted the cumulants with $\la\ra_\c$. 

Both $\mu_a$ and $\Delta_{ab}$ can be evaluated in the ZA by reading off $\bar f$ and $\langle f^2\rangle_{\mathrm c}$ from eq.~(\ref{2pfz_simple}) and eq.~(\ref{fbarZ}), using (\ref{fI}) and (\ref{ICcum}). One may think that the appropriate choice for the initial conditions is to keep $f(\eta_I)$ correct only to first order in $\delta(\eta_I)$, as in eq.~(\ref{2pfz_simple_expanded}), which is what \cite{Valageas:2003gm} uses. However, one should remember that the phase-space initial conditions for $\bar f$ and $\langle f^2\rangle_{\mathrm c}$ given by (\ref{fbarZ}) and (\ref{2pfz}) evaluated at $\eta_I$ should be kept consistent for $w$ up to $w\sim Dk$ (see discussion ca. eq.~(\ref{EPS_Z})), with $D$ evaluated today (or at the $\eta$ of interest) and not at $\eta_I$. Therefore, expanding eq.~(\ref{2pfz_simple}) to first order in $P_L$ is correct to $\mathcal{O}(\delta(\eta)^2)$ and not to $\mathcal{O}(\delta(\eta_I)^2)$. Thus, eq.~(\ref{2pfz_simple}) breaks down even at weakly non-linear scales, and one should use (\ref{2pfz}) instead.

One should note that it is only the overdensity and not the full one-particle distribution function which is a Gaussian random field. The full $f$ is in fact non-Gaussian and keeping the non-Gaussian parts will turn out to be crucial for the self-consistency of the formalism.

Indeed, in the ZA, the (higher $n$)-point functions are generated at higher order in $\delta(\eta_I)$. For example, the 3-pt function is generated at $\mathcal{O}(\delta(\eta_I)^4)$. However, following an analogous analysis as above, we can see that the effects from the initial (higher $n$)-point functions will actually kick in not at $\mathcal{O}(\delta(\eta_I)^4)$ but at $\mathcal{O}(\delta(\eta)^4)$ for weakly non-linear and nonlinear scales. 

However, in most of what follows we restrict ourselves to Gaussian initial conditions for $f$ in order to keep the formalism more transparent. Thus for now we will assume that the statistics of $f_I$ are entirely specified by $\mu_a$ and $\Delta_{ab}$. We will restore the non-Gaussianities in $f_I$ in Section~\ref{app:ngf}.

Note that Valageas \cite{Valageas:2003gm} expands $\Delta_{ab}$ to first order in $P_L$, as in eq.~(\ref{2pfz_simple_expanded}); and drops the non-Gaussianities of $f_I$. Therefore, his initial conditions for the 2-pt function of $f$ are only correct to first order in $\bm w$. This means that \cite{Valageas:2003gm} artificially smooths the initial conditions in the momentum direction. The expectation value $\int d^3p {\bm p}^n f_I\propto \partial^n/(\partial {\bm w}^n)f_I(w=0)$ is zero for $n>1$ under this approximation. Since the higher velocity moments vanish, the higher order velocity cumulants must be non-zero. Thus, the approximation in \cite{Valageas:2003gm} affecting the initial conditions, does not reduce to a pressureless fluid with vanishing shear stresses at $\eta_I$. Because of the structure of the equations, and the fact that Valageas works in a $P_L$ expansion in order to obtain a final result for the matter power spectrum, that assumption remains valid for all times and is not contained only to the initial conditions. 


\section{The extended BBGKY hierarchy and the response functions of the CDM}\label{sec:BBGKY}

In this section we review the derivation of the BBGKY hierarchy \cite{peebles}. Note that the BBGKY hierarchy as written in \cite{peebles} concerns only equal-time $n$-point functions. We drop that restriction in this paper. In the process we will also incorporate the initial conditions in the VP equation as done in (\ref{vp_stoch}), and the result will be what we refer to as the \textit{extended} BBGKY hierarchy for both the correlation and response functions.

One can try to solve the VP equation, (\ref{vp_stoch}), using a Born-type approximation, which schematically looks as follows:
\be\label{BornSeries}
f=L^{-1}f_{I}+L^{-1}Kf f=L^{-1}f_{I}+L^{-1}K(L^{-1}f_{I})(L^{-1}f_{I})+\cdots\ .
\ee
Note that we have written the series above so that the operator $K$ takes one argument on the left, and two on the right.
This series has a straightforward physical interpretation. Each term can be represented as a diagram, where several modes of the initial conditions are propagated forward in time by the linear operator, $L^{-1}$. Then, the modes interact one by one gravitationally through the 3-vertex, $K$, and the resulting mode is linearly propagated in time. It then interacts with other modes until one obtains the mode of interest ($f$ at a certain moment in time). To obtain the statistics of the phase-space density, one needs to multiply one or more $f$'s, as given by eq. (\ref{BornSeries}), and then take the ensemble average. One thus obtains that the $n$-point function is given by a series of integro-differential operators acting on the statistics of the initial conditions, $f_I$, given in (\ref{ICcum}). This approach quickly leads to a mindless proliferation of diagrams. One may try to systematically group different diagrams under certain meaningful physical quantities. One way to achieve this is by using the extended BBGKY hierarchy, which gives us the equations of motion for the correlation and response functions.

The first equation of the extended BBGKY hierarchy is obtained by taking the ensemble average of eq.~(\ref{vp_stoch}):
\be\label{proba2}
L_{ab}\bar f_b-K_{abc}\bar f_b \bar f_c-K_{abc}G_{bc}=\mu_a\ ,
\ee
where $\bar f_a\equiv \langle f_a\rangle$ and $G_{ab}\equiv \langle f_a f_b\rangle-\bar f_a\bar f_b= \langle f_a f_b\rangle_\c$ denotes the connected 2-point correlation function. For a homogeneous and isotropic universe, when the ensemble averaging is done without constraints (such as forcing the origin to be on top of a halo as in \cite{Ma:2003cq}, for example), we can see that the term $K_{abc}\bar f_b \bar f_c$ vanishes identically, since it is proportional to $\partial_{\bm x}\phi$ which vanishes for homogeneous $\bar f$ as per (\ref{jeans}).

Equation~(\ref{proba2}) is an equation for $\bar f_a$. To find $G_{ab}$ we need to multiply eq.~(\ref{vp_stoch}) by $f$ and \textit{then} take the ensemble average. The resulting equation is
\be\label{proba3}
(L_{ax}-2K_{axy}\bar f_y)G_{xb}-K_{axy}\langle f_xf_y f_b\rangle_\mathrm{c}=\langle f_{I,a} f_b\rangle_{\mathrm{c}}\ ,
\ee
where we used eq.~(\ref{proba2}). The three point function is given by multiplying (\ref{vp_stoch}) by $f^2$ and then taking the ensemble average. This yields:
\be\label{proba4}
(L_{ax}-2K_{axy}\bar f_y)\langle f_xf_bf_c\rangle_\mathrm{c}-2 K_{axy}G_{xb}G_{yc}-K_{axy}\langle f_xf_yf_bf_c\rangle_\c=\langle f_{I,a}f_bf_c \rangle_\mathrm{c}\ .
\ee
And in general, the $n$-point function is given by the cumulants up to the $(n+1)$-point function. 

Setting $f_I$ to zero in eq.s~(\ref{proba2},\ref{proba3},\ref{proba4}) we recover the standard BBGKY hierarchy which requires initial conditions to be set for the $n$-point functions (which is especially straightforward for Gaussian initial conditions). The presence of $f_I$ in those equations means that we have incorporated the initial conditions inside the equations at the cost of introducing cumulants of the type $\langle f_I f^m\rangle_\c$ which are needed in order to close the hierarchy.\footnote{This complication will allow us to transform the extended BBGKY hierarchy to the HH.} Those can be expressed via the response functions of the system. We will later show that for the CDM with Gaussian initial conditions:
\be\label{ffI}
\langle f^n f_{I,b_1}f_{I,b_2}\cdots f_{I,b_m}\rangle_\c=\left.\Delta_{b_1x_1}\Delta_{b_2x_2}\cdots\Delta_{b_mx_m}\frac{\delta^m\langle f^n\rangle_\c}{\delta\zeta_{x_1}\delta\zeta_{x_2}\cdots\delta\zeta_{x_m}}\right|_{\zeta=0}\ ,
\ee
where $\zeta$ is a constant-across-realizations forcing, which corresponds to changing the VP equation (\ref{vp_stoch}) by adding the forcing $\zeta$ to the stochastic $f_I$. Thus, $\delta^m\langle f^n\rangle_\c/\delta \zeta^m(\zeta=0)$ is the $n$-point response function of $m$-th order. From now on all derivatives with respect to $\zeta$ are meant to be evaluated at $\zeta=0$, even without stating that explicitly.

In order to obtain $\langle f_I f\rangle_{\mathrm{c}}$, which enters in eq.~(\ref{proba3}), we use (\ref{ffI}). Therefore, after that we need an equation for $\delta \bar f/\delta \zeta$. Taking the variation with respect to $\zeta$ of eq.~(\ref{proba2}) we obtain:
\be\label{fIzeta}
(L_{ax}-2K_{axy}\bar f_y)\frac{\delta \bar f_x}{\delta \zeta_b}-K_{axy}\frac{\delta G_{xy}}{\delta \zeta_b}=\frac{\delta \bar f_{I,a}}{\delta\zeta_b}=\d(\k_a-\k_b)\d(\o_a-\o_b)\d(\eta_a-\eta_b)\ ,
\ee
where in the last equality we used that adding a non-random $\zeta$ to $f_I$ corresponds identically to changing the mean of $f_I$. To remind the reader, all derivatives in $\zeta$ are to be evaluated at $\zeta=0$. Both the above equation and eq.~(\ref{proba4}) depend on $\delta G_{ab}/\delta \zeta$, which can be obtained by varying eq.~(\ref{proba3}). This in turn generates a term $\delta^2 \bar f/\delta\zeta^2$. And so on. 

Finding closure relations for the resulting hierarchy of equations for the correlation and response functions is a non-trivial task. One can simply set to zero all $n$-point correlation functions for $n$ above some cutoff. However, in the nonlinear regime this is inconsistent as the (higher $n$)-point correlation functions scale as products of (lower $n$)-point functions, which can be much bigger than 1 \cite{peebles}. Since nonlinear features are generated at small scales as soon as one starts evolving the VP system, such a truncation is unsatisfactory even at early times if we want to study the effective CDM dynamics.

Another option is to use some ansatz for the (higher $n$)-point functions to close the extended BBGKY hierarchy. However, such ansatzes are usually extracted only from observations or simulations (e.g. \cite{davis&peebles}).

In this paper, we investigate an alternative route. Each contraction entering in the extended BBGKY hierarchy between the operators $L$ and $K$, and the correlation and response functions can be represented by a diagram. We will argue that the HH corresponds to the extended BBGKY hierarchy with closure relations equivalent to dropping certain diagrams from the expressions for the correlation and response functions. The closure relations obtained in such a way are equivalent to dropping higher order vertices from the 1PI effective action. This in turn corresponds to performing infinite resummations of certain diagrams entering in the expressions for the correlation and response functions in a way which is easily generalizable to arbitrary order in the HH. We refer the reader to eq.s~(\ref{closure3pt}, \ref{closure2ptresponse}) derived below which give an example of a truncation to the extended BBGKY hierarchy obtained by a sharp truncation to the Helmholtz hierarchy.

\section{Statistical mechanics formulation of the Vlasov-Poisson equation}\label{sec:VP}
In statistical mechanics any analysis of the statistical properties of a system starts by writing down a partition function. In this section we rewrite the expectation value of any functional $F[f]$ as a path integral which will allow us to write down the partition function for the CDM. We do this by closely following \cite{Valageas:2003gm}.

\subsection{Path-integral formulation of the Vlasov-Poisson equation}\label{sec:piVP}

Following \cite{Valageas:2003gm}, we apply the Martin-Siggia-Rose (MSR) technique \cite{Martin:1973} \cite{Phythian:1977} to the stochastic differential equation (\ref{vp_stoch}) giving the evolution of the CDM. The method allows us to rewrite the ensemble average of any functional $F[f]$, where $f$ is the solution to a stochastic differential equation (in our case (\ref{vp_stoch})), into a path-integral form. Such a path-integral formulation will allow us to apply techniques borrowed from statistical mechanics to study structure formation.

The ensemble average of a functional, $F[f]$, can be written as:
\be\label{Fbegi}
\la F[f]\ra=\int \dD f_I \,F[f] \,e^{-\frac{1}{2}(f_{I,a}- \mu_a)\inv\Delta_{ab}(f_{I,b}- \mu_b)}\ ,
\ee
where any normalization factor is absorbed in $\dD$. The above equation is nothing but an average of $F[f]$ over the random initial conditions set by inflation, which are taken into account by the Gaussian kernel in the integral. The measure is defined as $\dD f_I\equiv \prod_{\bm x,\bm p} \mathrm{d} f_I(\bm x,\bm p)$ (up to a normalization factor), i.e. it goes over the field values on the initial Cauchy surface. In the above equation, $f$ is determined by $f_I$ using the equation of motion (\ref{vp_stoch}). We can rewrite (\ref{Fbegi}) as follows
\be\label{Fdelta}
\la F[f]\ra&=&\int \dD f_I\dD f \det\left[\L{a}{b}-K_{abc}f_c\right] F[f] e^{-\frac{1}{2}(f_{I,a}- \mu_a)\inv\Delta_{ab}(f_{I,b}- \mu_b)}\nonumber\\
&&\times \d\left(\L{a}{b}f_b-K_{abc}f_bf_c-f_{I,a}\right) \ ,
\ee
where the delta function imposes the equation of motion. The path-integral measure, $\dD f$, goes over all field configurations on all of phase-space and times later than $\eta_I$.  As shown in \cite{Valageas:2003gm}, the determinant reduces to an irrelevant constant, as long as we require that the Green's function for $\partial_{\eta_a}$, which enters in $\L{a}{b}$, is $\theta^0(\eta_a-\eta_b)$, where $\theta^a(x)$ is the Heaviside step function with the value at $x=0$ given by $\theta^a(0)=a$. As we will show later when we solve the NPRG flow equations, this choice for a Green's function will enforce causality\footnote{Note that $\theta^0(\eta_a-\eta_b)$ is not invertible, since it is triangular and has zeros on the diagonal $\eta_a=\eta_b$. In non-equilibrium statistical physics this is dealt with by working with discrete time. What one shows then is that actually there are purely imaginary numbers on the diagonal, which can be shown to be irrelevant. (e.g. \cite{Kamenev:2004})}.

The next step of the MSR method is to rewrite the delta function as follows
\be\label{Fchi}
\la F[f]\ra=\int \dD f_I\dD f \dD\chi F[f] e^{-\frac{1}{2}(f_{I,a}- \mu_a)\inv\Delta_{ab}(f_{I,b}- \mu_b)-\chi_a\left(\L{a}{b}f_b-K_{abc}f_bf_c-f_{I,a}\right)}\ ,
\ee
where $\chi$ is an \textit{imaginary} auxiliary field. We can now perform the Gaussian integral over $f_I$ and obtain
\be
\la F[f]\ra=\int \dD \phi F[f] e^{-S[\phi]}, \ \mathrm{where}\label{action} \\
S[\phi]\equiv\chi_a \L{a}{b}f_b-\chi_a  K_{abc}f_bf_c-\chi_a \mu_a-\frac{1}{2}\chi_a\Delta_{ab}\chi_b\label{S}\ ,
\ee
where we combined the fields $f$ and $\chi$ into the field vector $\phi=(f,\chi)$. Thus, the action $S[\phi]$ entirely encodes the statistics of the CDM evolution. Equation (\ref{action}) is the final result of this section.

\subsection{Response and correlation functions}

In this section we derive expressions for the response and correlation functions of $f$ based on the path-integral, eq. (\ref{action}). The final result will show that the role of the auxiliary field $\chi$ is to generate the response functions of CDM. 

Let us go back to the VP equation, (\ref{vp_stoch}). We can add a non-random (constant across realizations) forcing, $\zeta(\k,\o,\eta)$, on the right hand side of that equation. This modification can be taken into account in (\ref{Fbegi}) and (\ref{S}) by replacing $ \mu_a$ by $ \mu_a+\zeta_a$. The $n$-th functional derivative of $\la F[f] \ra$ with respect to $\zeta$ gives the response of $\la F[f] \ra$ under a change in the forcing $\zeta$. Using (\ref{action}) with $S\to S-\chi_a\zeta_a$ we find
\be\label{response}
\left.\frac{\delta^n \la F[f]\ra}{\delta \zeta_{a_1}\delta \zeta_{a_2}\cdots\delta \zeta_{a_n}}\right|_{\zeta=0}=\la F[f]\chi_{a_1}  \chi_{a_2}\cdots  \chi_{a_n}\ra\ .
\ee
For $F[f]=f_{b_1}f_{b_2}\cdots f_{b_m}$, we see that $\la F[f]\chi_{a_1}  \chi_{a_2}\cdots  \chi_{a_n}\ra$ reduces to the $m$-point response function of $n$-th order.

Causality requires that any effect from the non-random forcing $\zeta$ must propagate forward in time. This implies that for the bracket above to be non-zero, at least one of the $f$'s in the bracket must follow all $\zeta$'s, and hence all $\chi$'s as well:
\begin{flalign}\label{CC}
\lefteqn{\left.\frac{\delta^n \la f_{b_1}f_{b_2}\cdots f_{b_m}\ra}{\delta \zeta_{a_1}\delta \zeta_{a_2}\cdots\delta \zeta_{a_n}}\right|_{\zeta=0}=}\nonumber\\
&&=\la f_{b_1}f_{b_2}\cdots f_{b_m}\chi_{a_1}  \chi_{a_2}\cdots  \chi_{a_n}\ra\,\propto\, \theta^0\big[\max(\eta_{b_1},\cdots,\eta_{b_m})-\max(\eta_{a_1},\cdots,\eta_{a_n})\big]\ .
\end{flalign}
As we will later show, the choice for the value of $\theta(0)$ above is fixed by the requirement that the HH must not violate the above causality condition.

For $F[f]=1$, using $\la F[f]\ra=\la 1\ra=1$ we obtain
\be\label{ghost}
\left.\frac{\delta^n \la 1\ra}{\delta \zeta_{a_1}\delta \zeta_{a_2}\cdots\delta \zeta_{a_n}}\right|_{\zeta=0}=\la \chi_{a_1}  \chi_{a_2}\cdots  \chi_{a_n}\ra=0\ ,
\ee
which tells us that the auxiliary field itself is unobservable. From here one can show that
\be\label{responsecumulant}
\left.\frac{\delta \la f^m\ra_\c}{\delta\zeta^n}\right|_{\zeta=0}=\la f^m\chi^n\ra_\c\ ,
\ee
(shown empirically using Mathematica for a wide range of $m$ and $n$).

Let us define the Gibbs partition function (e.g. \cite{kardar}), $Z[j]$
\be\label{Z}
Z[j]\equiv \la e^{j_a\phi_a}\ra=\int \dD \phi  e^{-S[\phi]+j_a\phi_a}\ ,
\ee
where $a$ in $j_a$ and $\phi_a$ runs over field labels as well. Thus we have $j_a\phi_a=j_{f,a}f_a+j_{\chi,a}\chi_a$, where $j_{f,a}$ and $j_{\chi,a}$ are the external sources for $f$ and $\chi$, respectively. Taking functional derivatives with respect to $j_a$ and then setting $j_a$ to zero, we see that as usual $Z[j]$ generates the correlation and response functions of the system.

We are interested in the cumulants of $f$ (i.e. its connected correlation functions), and therefore we introduce yet another quantity, the Gibbs free energy\footnote{Note that in statistical mechanics the logarithm of the \textit{Gibbs} partition function (in the \textit{Gibbs} canonical ensemble) gives the Gibbs free energy (up to a proportionality constant); while the Helmholtz free energy is proportional to the logarithm of the partition function in the canonical ensemble \cite{kardar}. In quantum field theory (QFT) these definitions are usually swapped (e.g. \cite{peskin}). We adopt the usual statistical mechanics definition, unless otherwise noted.\label{foot:definitions}}, $W[j]$, given by 
\be\label{W}
W[j]\equiv\ln Z[j]\ .
\ee
 The Gibbs free energy generates the cumulants of $\phi$:
\be\label{Wcum}
\left.W_{;a_1a_2 \cdots a_n}\right|_{j=0}\equiv \left.\frac{\delta}{\delta j_{a_n}}\cdots\frac{\delta}{\delta j_{a_2}} \frac{\delta}{\delta j_{a_1}} W[j]\right|_{j=0}=\la\phi_{a_1}\phi_{a_{2}}\cdots\phi_{a_{n}}\ra_\c\ ,
\ee
where all subscripts after a semicolon denote functional derivatives as follows: For any functional $A$ of field $z$, we have $A[z]_{;\, a_1\cdots a_n}\equiv \delta^nA[z]/(\delta z_{a_n}\cdots\delta z_{a_1})$. Note that $\la \phi^n\ra$ is not always evaluated at $j=0$, as it sometimes denotes the ensemble average in the presence of the external source $j$. It should be clear from the context whether $\la \phi^n\ra$ is to be evaluated at $j=0$ or not.

From (\ref{ghost}) and (\ref{Wcum}), we conclude that $W[j]$ generates all connected correlation and response functions (see eq.~(\ref{responsecumulant})). As an illustration of the notation, let us consider the connected 2-point function for $f$, finding which is one of the goals of this paper. It can be obtained from $W_{;ab}$, which can be written in  block form as follows\footnote{For a 2-index operator, such as $W_{;ab}$, we use the matrix notation to explicitly show the dependence on the field components of $\phi$ (or their respective external sources), i.e. $f$ and $\chi$. The (1,1) component in a matrix corresponds to the ($f_af_b$) component of the operator; (2,2) -- corresponds to the ($\chi_a\chi_b$) component; (1,2) -- to the $(f_a\chi_b)$ component; and (2,1) -- to the $(\chi_af_b)$ component. See for example eq.~(\ref{Wtable}).}:
\be\label{Wtable}
\renewcommand{\arraystretch}{1.5}
\left.W_{;ab}\right|_{j=0}=\left.\left(
\begin{array}{cc}
\frac{\delta}{\delta j_{f,b}} \frac{\delta}{\delta j_{f,a}}  & \frac{\delta}{\delta j_{\chi,b}} \frac{\delta}{\delta j_{f,a}}   \\
\frac{\delta}{\delta j_{f,b}} \frac{\delta}{\delta j_{\chi,a}}  & \frac{\delta}{\delta j_{\chi,b}} \frac{\delta}{\delta j_{\chi,a}} 
\end{array}
\right)W[j]\right|_{j=0}=\left(
\begin{array}{cc}
\la f_a f_b\ra_\c & \frac{\delta\bar f_a}{\delta\zeta_b}  \\
 \frac{\delta\bar f_b}{\delta\zeta_a} & 0
\end{array}
\right),
\ee
where we used (\ref{response}) and (\ref{ghost}). 
Thus, we converted the problem of finding $\la f_af_b\ra_\c$ to a problem of finding\footnote{The matrix $W_{;ab}$ shown above gives exactly the three non-equilibrium Green's functions which appear in the Keldysh formalism for studying non-equilibrium statistical systems (see e.g. \cite{Kamenev:2004} for a discussion of the relation between the Keldysh formalism and the MSR technique).} $W_{;ab}$. 

As an aside, let us derive eq.~(\ref{ffI}) which gives the cumulant $\la f^m f_I^n\ra_\c$ entering in the extended BBGKY hierarchy discussed in Section \ref{sec:BBGKY}. In analogy with \cite{Valageas:2007ge}, we express $f_I$ from the VP equation and find
\be\label{tempffI}
\la f^m f_I^n\ra=\int \dD \chi \dD f \,f^m \left(-\frac{\delta}{\delta \chi}+\mu+\Delta\cdot\chi\right)^n  e^{-S[\phi]}\ ,
\ee
where $(\Delta\cdot\chi)_{a}\equiv\Delta_{ax}\chi_x$. Integrating by parts the above equation $n$ times and taking the connected part of the resulting expression, we recover\footnote{We showed that eq.~(\ref{ffI}) can be obtained in the above described way from eq.~(\ref{tempffI}) using Mathematica for a wide range of $m$ and $n$.} eq.~(\ref{ffI}). Note that eq.~(\ref{ffI}) is exact even in the nonlinear regime. As the above derivation shows, this is due entirely to the fact that we deal with ensemble averages. In analogy with (\ref{ffI}) we can express the ensemble average of the product of any functional $F$ of $f$ with $f_I^m$ as
\be\label{ffIF}
\langle F[f] f_{I,b_1}f_{I,b_2}\cdots f_{I,b_m}\rangle_\c=\left.\Delta_{b_1x_1}\Delta_{b_2x_2}\cdots\Delta_{b_mx_m}\frac{\delta^m\langle F[f]\rangle_\c}{\delta\zeta_{x_1}\delta\zeta_{x_2}\cdots\delta\zeta_{x_m}}\right|_{\zeta=0}\ .
\ee

By analogy, we can calculate $\delta \bar f_I/\delta \zeta$ entering in eq.~(\ref{fIzeta}):
\be
\left.\frac{\delta \bar f_{I,a}}{\delta\zeta_b}\right|_{\zeta=0}=\int \dD \chi \dD f \,\chi_b \left(-\frac{\delta}{\delta \chi_a}+\mu_a+\Delta_{ax}\chi_x\right)  e^{-S[\phi]}=\delta_{ab}\ ,
\ee
where the last equation is obtained after integration by parts. The above result confirms the result from the discussion after eq.~(\ref{fIzeta}).

\subsection{Summary}
Following \cite{Valageas:2003gm}, in this section we wrote the ensemble average of any functional of $f$, $F[f]$, in a path integral form using the MSR technique applied to the VP equation. We started off by writing the ensemble average of $F[f]$ as an integral over the initial phase-space configurations of $f$, weighted by the Gaussian random initial conditions set by inflation.

The integral can be rewritten as an integral over all phase-space configurations of the 1-particle density function of the CDM. We showed that the path integral is over the exponential of an action, which is the classical action governing the CDM dynamics in the presence of the stochastic initial conditions -- a result which is standard in the MSR method, and which was first derived for the CDM by \cite{Valageas:2003gm}. We were able to write down that path integral at the expense of introducing an auxiliary field, $\chi$. This field is unobservable; it generates the response functions of the CDM; and by construction it enforces the VP equation of motion for $f$ for each realization of the initial conditions.

Using the path integral formalism, we obtained the Gibbs partition function, which generates the correlation functions of the fields, $f$ and $\chi$. The logarithm of the partition function is the Gibbs free energy, which generates the cumulants of the fields. Therefore, the equations of motion for the derivatives of the Gibbs free energy must be exactly given by the extended BBGKY hierarchy.

\section{NPRG flow equations}\label{sec:NPRG}
In this section we will Legendre transform the extended BBGKY hierarchy to obtain the HH. We will start by writing down an exact equation for $W[j]$ -- the generator of all correlation and response functions of the CDM phase-space distribution function. We use the Non-Perturbative Renormalization Group (NPRG) flow based on the effective average action method \cite{Wetterich:1992yh} and its variants for obtaining $W$, which provide an exact non-perturbatively correct equation for $W$ \cite{Wetterich:1992yh}. Our discussion of the NPRG flow is based on \cite{Berges:2000ew}.

In the next section we will be able to integrate the NPRG flow equations and will end up with the Helmholtz hierarchy. We will argue that the Helmholtz hierarchy is equivalent to the extended BBGKY hierarchy discussed in Section \ref{sec:BBGKY} after a Legendre transformation. Thus, we will argue that a truncation of the NPRG (or equivalently, the Helmholtz) hierarchy ameliorates the closure problem of the BBGKY hierarchy.

\subsection{Cutoff}

The idea behind the Renormalization Group (RG) flow is to obtain an action which describes the ``interesting'' degrees of freedom (e.g. the low-$k$ modes), $\phi_<$, after one averages over (``integrates out'') the ``uninteresting'' degrees of freedom (e.g. the high-$k$ modes), $\phi_>$. Let us parametrize the boundary between interesting and non-interesting degrees of freedom by some cutoff parameter $\lambda$. The resulting action for the interesting degrees of freedom, $S_\lambda$, can be easily obtained from the partition function $Z$ (\ref{Z}) if the functional integration is performed only over $\phi_>$. Thus we obtain
\be
\exp\{-S_\l\}\equiv \int\dD\phi_> \exp\{-S[\phi_<,\phi_>]+j_>\phi_>\}\ ,
\ee
where we used that $\phi$ in (\ref{Z}) runs over $\phi_<$ and $\phi_>$; $j_<$ and $j_>$ are the external sources corresponding to $\phi_<$ and $\phi_>$, respectively. The flow of $S_\l$ with $\l$ is contained in Wilson's approach \cite{Wilson:1973jj} to renormalization theory. Once all degrees of freedom are integrated over we can see from (\ref{W}) that $S_\l\to -W$. Thus, $S_\l$ interpolates between $S$ and $-W$. This is the basic tenet of renormalization theory: to find a coarse-grained (i.e. high-$k$ modes are integrated over for a cutoff in $k$) action $S_\l$ governing the behavior of the ``interesting'' modes, starting from the ``bare'' action, $S$.

The transition between integrated ($\phi_>$) and unintegrated ($\phi_<$) degrees of freedom can be made smooth if one introduces ``incomplete integration'' \cite{Wilson:1973jj}. To understand how incomplete (or ``partial'') integration can be achieved, we give an example with a one-dimensional integral over some function $f(x)$:
\be
f_\l(x)=\int dy f(y)r_\l(x,y)\ , \ \ \mathrm{with}\ \ r_\l(x,y)=e^{-\l (x-y)^2}\ ,
\ee
with $r_\l(x,y)$ called the ``cutoff'' function (compare the above equation with eq.~(\ref{GammaPath})). We can see that the ``flow'' of $f_\l$ satisfies the following boundary conditions
\be
f_{\l\to \infty}(x)=\mathrm{constant}\times f(x)\ , \ \ \mathrm{and} \ \ f_{\l\to 0^+}(x)=\int dy f(y)\ .
\ee
Therefore, $f_\l$ interpolates between $f$ and its integral. Thus, $r_\lambda$ provides exactly the smooth transition we need to go from $\phi_<$ to $\phi_>$. 

Following that prescription, the NPRG flow equations for the CDM dynamics can be derived after one modifies the partition function (\ref{Z}) by introducing a cutoff function, $\R^\l_{ab}$, depending on a cutoff parameter, $\lambda$:
\be\label{Zcut}
Z_\lambda[j]\equiv \int \dD \phi  e^{-S[\phi]+j_a\phi_a-\frac{1}{2}\phi_a\R^\lambda_{ab}\phi_b}\ .
\ee
It is important to note that the normalization constant entering in $\dD\phi$ above is chosen to be $\lambda$-independent, i.e. it equals $1/Z[0]$. We impose this requirement for simplicity, so that under differentiation with respect to $\lambda$, the normalization constant does not produce any extra terms. 

Broken into field components, in our case the cutoff can be written as
\be
\renewcommand{\arraystretch}{1.5}
\R^\lambda_{ab}=\R^\lambda_a \left(
\begin{array}{cc}
 \delta_{a,-b} & 0 \\
 0 &\delta_{a,-b}
\end{array}
\right)\ ,
\ee
where $\R^\lambda_a$ is positive and changes between zero and infinity\footnote{Another requirement for $\R^\lambda_a$ is to rise faster than any power of the fields, so that in the end we are not left with incompletely integrated degrees of freedom. However, we do not bother with these details, since in the end we take the limit where $\R^\lambda_a$ is a sharp cutoff switching between zero and infinity.}. Note that $\delta_{a,-b}$ in the $(f_af_b)$ and  $(\chi_a\chi_b)$ components of $\R^\lambda_{ab}$ ensures that the cutoff term in $Z_\lambda[j]$ has a negative overall sign.

In the exact NPRG flow of Wetterich \cite{Wetterich:1992yh}, the cutoff function $\R^\lambda_{ab}$ is chosen to be a momentum\footnote{Note that in the context of the RG flow, by ``momentum'' we mean the wave-vector $\bm k$, and not the physical momentum of the CDM particles.\label{foot:momentum}} cutoff as follows: For $\lambda\gg k_a$ the cutoff diverges, $\R^\lambda_a\to\infty$, while for $\lambda\ll k_a$, the cutoff is set to zero. However, in what follows we keep the nature of the cutoff $\R^\lambda_{ab}$ and the bare action, $S[\phi]$, completely arbitrary.

\subsection{Effective action (Helmholtz free energy)}\label{subs:eff_act}

The NPRG flow equation is easiest to write down for the Helmholtz free energy, $\Gamma[\vf]$ (defined below), which is related to the Gibbs free energy via a Legendre transformation. In the absence of a cutoff, the field $\vf$ is defined as 
\be
\vf_a\equiv \frac{\delta W[j]}{\delta j_a}\ .
\ee
Therefore, $\vf=\la \phi\ra$ is the average (``classical'') field evaluated in the presence of an external source, $j$ \cite{peskin}.

Later (cf. eq.~(\ref{EqOfM})) we will see that in the absence of the cutoff, $\vf_a$ obeys $\delta\Gamma[\vf]/\delta\vf=j=0$, i.e. one has to extremize $\Gamma$ to obtain the dynamics of the classical fields. In quantum field theory, finding the extremum of $\Gamma$ corresponds to finding the equilibrium field configuration while including all effects of quantum fluctuations and at the same time setting the external sources to zero. In the context of a classical magnetic system, our restriction ($j=0$) implies finding the thermodynamic configuration under zero external magnetic field, but including the effect of thermal fluctuations. In the context of CDM dynamics, varying $\Gamma$ corresponds to changes to the system while including all effects of the stochastic initial conditions and setting any non-random forcings to zero. The last statement can be seen from the fact that any $j_\chi$ can be absorbed in the initial condition $\mu$, in a similar manner as we did for the non-random forcing, $\zeta_a$, discussed ca. eq.~(\ref{response}). One can view $\Gamma$ as an action, and the corresponding equations of motion (cf. eq.~(\ref{j}) or eq.~(\ref{EqOfM})) are the ones governing the dynamics of the classical field, $\vf$, hence the alternative name for $\Gamma$ -- ``effective action''. 

In the presence of the cutoff, all thermodynamic quantities acquire a dependence on the cutoff parameter, which we put as a subscript explicitly only for $Z$, $W$ and $\Gamma$ for brevity. The Gibbs free energy is given by 
\be\label{Wcut}
W_\l=\ln Z_\l\ ,
\ee
 while $\Gamma_\l$ is given below. The classical fields and the external sources also depend on $\l$, and unless otherwise specified, from now on $\vf$ and $j$ will denote the $\l$-dependent quantities, $\vf_\l$ and $j_\l$, respectively. We require:
\be\label{vfl}
\vf_{a}\equiv(\partial_{j_{a}}W_\l[j_{a}])_{\l} \ \ \hbox{in the presence of a cutoff $\R^\lambda_{ab}$}\ ,
\ee
where a subscript after brackets enclosing a partial derivative indicate a quantity held fixed when doing the differentiation. 

It will prove useful (see \cite{Berges:2000ew} and our Section~\ref{sec:IC}) to modify the usual relation between $\Gamma_\lambda$ and $W_\lambda$ as follows:
\be\label{Gamma}
\Gamma_\lambda[\vf_{a}]+W_\lambda[j_{a}]=j_{a}\vf_{a}-\frac{1}{2}\vf_{a}\R^\lambda_{ab}\vf_{b}\ .
\ee
As $\lambda$ changes, the value of $\R^\lambda_a$ for fixed $(\k_a,\o_a,\eta_a)$ starts at infinity and then is reduced to zero. Thus, $\Gamma_\lambda$ and $W_\lambda$ start at some initial conditions in functional space when $\R^\lambda_a$ diverges for all $(\k_a,\o_a,\eta_a)$ which are covered by the path integral in eq.~(\ref{Fdelta}) (i.e. $\eta_a>\eta_I$), and ``flow'' with $\lambda$ until they reach the true $\Gamma$ and $W$ when $\R^\lambda_a\to 0$ for all $(\k_a,\o_a,\eta_a)$. This flow in functional space is the functional RG flow generated by NPRG flow methods.

\subsection{Deriving the NPRG flow equation}

Before deriving the NPRG flow equation let us derive several helpful relations between the derivatives of the free energies. In the presence of a cutoff, derivatives denoted by a semicolon are taken at fixed $\l$, i.e. $\Gamma_{\l;a}\equiv(\partial_{\vf_{a}}\Gamma_{\l})_{\l}$.

Taking the derivative of (\ref{Gamma}) with respect to $\vf$ keeping $\l$ fixed we obtain \cite{Berges:2000ew}
\be\label{j}
\Gamma_{\l;a}=j_{a}-\R^\lambda_{ab}\vf_{b}\ ,
\ee
where we used 
$$
\bigg(\partial_{\vf} W_\l\big[j(\vf)\big]\bigg)_\l=(\partial_{\vf} j)_\l(\partial_{j} W_\l)_\l=(\partial_{\vf} j)_\l\, \vf\ .
$$ 
Note that equation (\ref{j}) gives the equation of motion for the classical field $\vf$. 

The NPRG flow is most easily derived by looking at $(\partial_\l \Gamma_\l)_{\vf}$. Taking the derivative of (\ref{Gamma}) in $\l$ at fixed $\vf$ we obtain
\be\label{prelimNPRG1}
(\partial_\lambda \Gamma_\l)_{\vf}+(\partial_\l W_\l)_{j}+\frac{1}{2}\vf_{a}  \dot\R^\l_{ab}\vf_{b}=0\ ,
\ee
where we used (\ref{vfl}) combined with
\be\label{W_fi}
(\partial_\l W_\l)_\vf=(\partial_\l W_\l)_j+(\partial_\l j)_\vf (\partial_j W_\l)_\l\ .
\ee
In eq.~(\ref{prelimNPRG1}), we denoted $\dot\R^\l_{ab}\equiv\partial_\l \R^\l_{ab}$ (This is the only exception to the rule that a dot represents a partial derivative with respect to conformal time). The derivative $(\partial_\l W_\l)_{j}$ can be obtained by working with (\ref{Zcut}) and (\ref{Wcut}). One obtains
\be\label{WNPRG}
(\partial_\l W_\l)_{j}=-\frac{1}{2}\dot\R^\l_{ab}\left[W_{\l;ba}+\vf_b\vf_a\right]\ .
\ee
Combining (\ref{prelimNPRG1}) and (\ref{WNPRG}) we end up with
\be\label{prelimNPRG2}
(\partial_\lambda \Gamma_\l)_{\vf}=\frac{1}{2}\dot\R^\l_{ab}W_{\l;ba}\ .
\ee

Now let us express $W_{\l;ba}$ using $\Gamma_\l$. Taking the derivative with respect to $\vf$ at constant $\l$ of (\ref{j}) and using (\ref{vfl}) we obtain
\be\label{inverseW2}
&&\delta_{ab}=(\Gamma_{\l;ax}+\R^\l_{ax})W_{\l,xb}=W_{\l,ax}(\Gamma_{\l;xb}+\R^\l_{xb})\ , \ \ \hbox{and therefore}\nonumber\\
&&W_{\l;ab}=\inv{\left(\Gamma_{\l}^{(2)}+\R^\l\right)}_{ab}\label{Winv}\ ,
\ee
where a superscript $(n)$ denotes the $n$-th functional derivative; and $\R^\l$ is a shorthand for $\R^\l_{xy}$. 

Combining (\ref{prelimNPRG2}) with (\ref{Winv}) we end up with \cite{Wetterich:1992yh}:
\be\label{NPRG}
(\partial_\lambda \Gamma_\l)_{\vf}=\frac{1}{2}\dot\R^\l_{ab}\inv{\left(\Gamma_{\l}^{(2)}+\R^\l\right)}_{ba}\ .
\ee

Equation (\ref{NPRG}) is the final result of this section. It is a NPRG flow equation as derived by \cite{Wetterich:1992yh} in the framework of the effective average action method. Taking functional derivatives of the NPRG equation (\ref{NPRG}) with respect to the field $\vf$ at fixed $\l$ one obtains an infinite hierarchy of equations for the functional derivatives $\Gamma^{(n)}_\l$ of $\Gamma_\l$. Given $\Gamma^{(n)}_\l$ one can derive $W^{(n)}_\l$, which is our final goal when evaluated at the final $\l$.

As an example, the functional flow for $\Gamma^{(1)}_\l$ is given by:
\be
(\partial_\l \Gamma_{\l;a})_{\vf}=-\frac{1}{2}\Gamma_{\lambda;axy}\left\{W_{\lambda;xv}\dot\R^\l_{vw}W_{\l;wy}\right\}\ ,
\ee
where $W_\l^{(2)}$ is always to be understood as given by (\ref{Winv}). Neglecting
 $\Gamma^{(4)}_\l$, the flow equation for $\Gamma^{(2)}_\l$ is given by
\be\label{2ndhier}
(\partial_\l \Gamma_{\l;ab})_\vf=\Gamma_{\lambda;awx}\Gamma_{\lambda;byz}W_{\l;wy}\left\{W_{\lambda;xu}\dot\R^\l_{uv}W_{\l;vz}\right\}+\mathcal{O}(\Gamma^{(4)})\ .
\ee
And in general the flow of $\Gamma^{(n)}_\l$ is determined by all functional derivatives of $\Gamma_\l$ up to $\Gamma^{(n+2)}_\l$.

Note that in principle one can work with the NPRG flow equation written for $W_\l$ (\ref{WNPRG}) which in a full treatment is equivalent to the NPRG flow equation for $\Gamma_\l$. However, $W_\l$ and $\Gamma_\l$ are related via a nonlinear transformation (\ref{Gamma}). Thus, the functional hierarchies for $\Gamma_\l^{(n)}$ and $W_\l^{(n)}$ do not result in equivalent dynamics under a truncation at one and the same order $n$. From QFT we know that $W$ generates all connected diagrams, while $\Gamma$ generates all 1-particle irreducible (1PI) diagrams\footnote{One can extend the above analysis and consider the NPRG flow of $N$-particle irreducible effective actions (e.g. \cite{Pawlowski:2005xe}). However, they generate further complications, and therefore we do not consider them here.} (i.e. those connected diagrams which cannot be disconnected by a removal of a single internal line). Thus, one can expect that a truncation of the $\Gamma_\l^{(n)}$ hierarchy will give more accurate results than a truncation of the $W_\l^{(n)}$ hierarchy at the same order in the functional expansion.

\subsection{Initial conditions for the NPRG flow}\label{sec:IC}

The initial conditions for the NPRG flow are set at the initial value of $\l=\l_i$ when $\R^{\l_i}_a$ diverges for all $(\k_a,\o_a,\eta_a>\eta_I)$. Using (\ref{Zcut}), (\ref{Wcut}) and (\ref{Gamma}), we can find the initial conditions for $\Gamma_{\l_i}$ by first writing:
\begin{eqnarray}\label{GammaPI}
\exp\left(-\Gamma_\l\right)&=&\exp\left\{
W_\l-j_{a}\vf_{a}+\frac{1}{2}\vf_{a}\R^\l_{ab}\vf_{b}
\right\}\\\nonumber
&=&
\int\dD\phi\exp\left[-S[\phi]-\frac{1}{2}\phi_{a}\R^\l_{ab}\phi_{b}+j_{a}\phi_{a}-j_{a}\vf_{a}+\frac{1}{2}\vf_{a}\R^\l_{ab}\vf_{b}\right]\ .
\end{eqnarray}
We can substitute $\phi\to\phi+\vf$ in the above equation  and combine with (\ref{j}) to obtain
\be\label{GammaPath}
\exp\left(-\Gamma_\l\right)=\int\dD\phi\exp\left[-S[\phi+\vf]+\phi_a\Gamma_{\l;a}-\frac{1}{2}\phi_{a}\R^\l_{ab}\phi_{b}\right]\ .
\ee
When $\R^\l$ diverges for all $(\k_a,\o_a,\eta_a>\eta_I)$, $\exp[-\R^\l\phi^2]$ becomes a delta function, and therefore we obtain the initial conditions for the NPRG flow \cite{Berges:2000ew}:
\be\label{IC}
\Gamma_{\l_i}[\vf_a]=S[\vf_a]
\ee
up to an irrelevant additive constant. The above equation holds for $\eta_a=\eta_I$ as well. This can be seen from the fact that at $\l=\l_i$ for $\eta_a=\eta_I$ the cutoff diverges for all $\eta> \eta_I$; and the fact that the path integral does not go over $\eta_I$ and therefore for that $\eta_a$ we have $\phi\to\vf$ in (\ref{GammaPI}).  The above simple result, eq.~(\ref{IC}), explains the modification that we made in eq.~(\ref{Gamma}) to the usual definition of $\Gamma$.

\subsection{Summary}\label{sec:NPRGsum}
In this section we reviewed the formalism behind the NPRG flow equation (\ref{NPRG}) obtained first in \cite{Wetterich:1992yh}. The NPRG flow describes the exact, nonperturbatively correct evolution of the effective action, $\Gamma_\l$, in functional space. The initial condition for the flow with $\l$ is the ``bare'' action, i.e. $\Gamma_{\l_i}[\vf]=S[\vf]$ (in the case of CDM it is given in eq.~(\ref{S})). When the cutoff $\lambda$ reaches a final value, $\l_f$, for which the cutoff function, $\R^\lambda$, vanishes for all $(\k_a,\o_a,\eta_a)$, we recover the full effective action. Thus, the flow of $\Gamma_\l$ interpolates between the bare action, $S$, and the full effective action, $\Gamma$. Note that the equations of motion for the functional derivatives, $\Gamma^{(n)}$, are precisely the Helmholtz hierarchy (We postpone further discussion of the HH until the next section). 

The NPRG flow can be written as an infinite hierarchy of equations governing the evolution of the functional derivatives $\Gamma^{(n)}_\l$ of $\Gamma_\l$. Truncating the hierarchy at some order $n$ in the functional expansion provides us with solutions which are functional approximations to the true $\Gamma^{(n)}$. Once we have $\Gamma^{(n)}$ up to some $n$, we can obtain $W^{(n)}$ (by differentiating eq.~(\ref{Winv}) and using (\ref{vfl})) which contains the correlation and response functions of CDM.

The final result for the full effective action does not depend on $\R^\l$ (as long as it diverges/decays fast enough in its two regimes), but the choice of $\R^\l$ does affect the quality of the approximation once a truncation of the hierarchy is enforced. Thus, certain optimization schemes have been developed to deal with the problem \cite{Pawlowski:2005xe}. Moreover, certain choices for $\R^\l$, such as a sharp cutoff, allow for the analytical integration of the flow. In what follows we concentrate on using a sharp cutoff, which will allow us to avoid resorting to numerical integrations as much as possible.

Following the original idea of \cite{Wilson:1973jj}, the standard choice for a cutoff is a cutoff in momentum space (see footnote \ref{foot:momentum}). The utility of a momentum cutoff is that for $\lambda$ between $\l_i$ and $\l_f$, it provides us with effective equations of motion for the coarse-grained field (i.e. after the high-$k$ modes have been integrated out) derived from $\Gamma_\l$. Thus, the high-$k$ physics influences the low-$k$ physics by modifying the low-$k$ equations of motion. In the context of CDM, this results in an effective sound speed and viscosity for the long-wavelength cosmological fluid \cite{Baumann:2010tm}; or after introducing certain constraints, this may result in effective Fokker-Planck terms in the effective equations of motion (see e.g. \cite{Ma:2003cq} for an approximate effective equation for $f$ describing an ensemble averaged halo profile). However, the result of \cite{Baumann:2010tm} is a result for the effective equation of motion for $f$ in a single realization. Extracting from that the nonlinear evolution of the statistical properties of CDM involves either a numerical simulation, or an analysis similar to the one presented here. Moreover, the parameters of the effective equation of motion for CDM still have to be extracted from numerical simulations for example \cite{Baumann:2010tm}. One can try to obtain those parameters from the evolution equations for the correlation and response functions obtained by the brute-force application of a NPRG flow. However, under a cutoff in $\bm{k}$ the resulting equations obtained from the NPRG flow are not manifestly causal (see footnote \ref{foot:equil}). That is not a problem when the method is applied to systems in equilibrium, since they are invariant under time translations. However, in the non-equilibrium setting of CDM dynamics the interpretation of these results becomes severely convoluted.

Thus, alternative approaches have been developed for non-equilibrium systems based on the NPRG flow, with a cutoff function not in momentum but in time \cite{Gasenzer:2007}. In the context of CDM, this method works by allowing the coupling of modes to take place up to a cutoff in time. This cutoff is moved from the initial time to the time, which one is interested in. Such a choice for the cutoff function generates a manifestly causal NPRG flow. The end result of such a procedure are time evolution equations for $\Gamma^{(n)}$ --- the Helmholtz hierarchy.

In the next section we concentrate on the NPRG flow with a sharp temporal cutoff which will allow us to integrate the NPRG flow, and obtain the HH.

\section{Integrating the NPRG equations for the CDM. The Helmholtz Hierarchy.}\label{sec:SOL}

In this section we are going to integrate the NPRG flow equations. We call the resulting hierarchy the ``Helmholtz hierarchy'', since it is a hierarchy for the functional derivatives of the Helmholtz free energy, $\Gamma$. As discussed in Section \ref{sec:NPRGsum}, unlike standard NPRG methods, we are going to use a temporal cutoff as proposed in \cite{Gasenzer:2007}. The temporal cutoff can be interpreted as follows. The nonlinear evolution couples pairs of modes at different times (compare with the discussion ca. eq.~(\ref{BornSeries})). In the NPRG with the temporal cutoff, the modes are allowed to interact only until the cutoff time. As the cutoff is moved forward in time, the gravitational interactions are allowed to take place until later times. The discussion in this section closely follows \cite{Gasenzer:2007}, since the path integral formulation of the CDM evolution is trivially related to the non-equilibrium formalism developed in \cite{Gasenzer:2007} by a standard change of variables, usually referred to as Keldysh rotation.

We spend a big part of this section to show that the sharp temporal cutoff results in a hierarchy which is consistent with causality and the fact that the $\chi$ field must be unobservable. These results must hold in the full nonperturbative treatment, but we perform that analysis to show that they also hold for our choice of cutoff after a truncation of the NPRG hierarchy. 

\subsection{Causality of the flow I}\label{sec:causalityI}
Having introduced the NPRG flow equation in Section \ref{sec:NPRG}, our starting point is to write down our choice for a cutoff function:
\be\label{cutoff}
\R^\t_a=\left\{ 
\begin{array}{l l}
  \infty & \quad \text{if $\eta_a>\tau$}\\
  0 & \quad \text{otherwise}\\
\end{array} \right.
\ee
such that $\R^\t_a\propto  \theta^0(\eta_a-\t)$, which matches the choice made in \cite{Gasenzer:2007}. Moreover, we adopted the notation $\t$ for the temporal cutoff from \cite{Gasenzer:2007} in order to highlight that our analysis from now on will be based on the sharp temporal cutoff, and will no longer be valid for arbitrary cutoff functions.

The utility of the choice (\ref{cutoff}) becomes apparent once we consider $W_{\t;a_1a_2\cdots a_m}$ which is a sum of products of $\langle \phi^n\rangle$ ($m\geq n$), with the expectation values taken in the presence of a fixed $j$:
\begin{eqnarray} \label{causal_n}
\langle \phi^n\rangle&=&\frac{ \int \dD \phi \, \phi_{a_1}\cdots\phi_{a_n} e^{-S[\phi]+j_a\phi_a-\frac{1}{2}\phi_a\R^\t_{ab}\phi_b}}{\int \dD \phi  e^{-S[\phi]+j_a\phi_a-\frac{1}{2}\phi_a\R^\t_{ab}\phi_b}}\\&=&
\frac{ \int \dD \phi \, \phi_{a_1}\cdots\phi_{a_n} e^{-S[\phi]+\Gamma_{\tau;a}\phi_a-\frac{1}{2}(\phi_a-\vf_a)\R^\t_{ab}(\phi_b-\vf_b)}}{\int \dD \phi  e^{-S[\phi]+\Gamma_{\tau;a}\phi_a-\frac{1}{2}(\phi_a-\vf_a)\R^\t_{ab}(\phi_b-\vf_b)}}\nonumber\\
\ \ & =& \vf_{a_1}\cdots\vf_{a_r}
\frac{ \int \dD \phi \, \phi_{a_{r+1}}\cdots\phi_{a_n} e^{-S[\phi]+j_a\vf_a}}{\int \dD \phi  e^{-S[\phi]+j_a\vf_a}}=\vf_{a_1}\cdots\vf_{a_r}\langle \phi_{a_{r+1}}\cdots\phi_{a_n} \rangle\ .\nonumber
\end{eqnarray} 
The last line of eq.~(\ref{causal_n}) is written for the case when $r$ of the $\vf$'s are inside the cutoff\footnote{By $\vf_a$ being ``inside'' or ``outside'' the cutoff we mean that $\R^\l_a\to\infty$ or 0, respectively.}. In (\ref{causal_n}) we used (\ref{j}).

From eq.~(\ref{causal_n}) one can see that as long as $r>0$, the \textit{connected} $m$-point function (for $m>1$) vanishes, and thus $W^{(m)}=0$. Therefore, if $\R^{\t>\eta_I}_a\to \infty$ for any of the $\eta_i$ ($i=1,\dots,m$), $W_{\t>\eta_I;a_1a_2\cdots a_m}$ vanishes. Since the bare action for the CDM contains only equal time operators (see eq.~(\ref{S})), as long as $\t$ is larger than $\max(\eta_{a_1},\cdots,\eta_{a_m})$, $\langle \phi^m \rangle$ ($m\geq0$)  is $\t$-independent. Thus, we can see that for $n=1$:
\be
\vf_{\t\geq \eta_a,a}=\vf_{\t=\eta_a,a}\ ,
\ee
and for $n>1$:
\be\label{Wt}
W_{\t>\eta_I;a_1a_2\cdots a_n}&=&W_{\t_{a_1\cdots a_n};a_1a_2\cdots a_n} \theta^1(\t-\t_{a_1\cdots a_n})\\\nonumber
&=&W_{;a_1a_2\cdots a_n} \theta^1(\t-\t_{a_1\cdots a_n})\ , \ \ \mathrm{with}\ \ \t_{a_1\cdots a_n}\equiv\max(\eta_{a_1},\cdots,\eta_{a_n})\ ,
\ee
where we have $\theta^1(\t-\t_{a_1\cdots a_n})=\prod_{i=1}^n \theta^1(\t-\eta_{a_i})$. The above equation trivially implies
\be\label{dWt}
(\partial_\t W_{\t>\eta_I;a_1a_2\cdots a_n})_j=W_{\t_{a_1\cdots a_n};a_1a_2\cdots a_n}\d(\t-\t_{a_1\cdots a_n}) \ \ \hbox{for} \ \ n\geq2\ ,
\ee
where one must be careful when integrating (\ref{dWt}) to use the correct step function as given in (\ref{Wt}).
Thus, the flow of $W^{(n)}_\t$ freezes after $\t$ exceeds the maximum of the time arguments of the response and correlation functions entering in $W^{(n)}_\t$ as required by causality. 

The flow for $\Gamma^{(n)}_\t$ must have the same causal property, i.e.
\be\label{Gamma_causal}
(\partial_\t\Gamma_{\t;a_1,a_2,\cdots,a_n})_\vf=0\ , \ \ \hbox{for}\ \ \t\geq \t_{a_1\cdots a_n}\ .
\ee
Combining (\ref{Winv}) and (\ref{Wt}) we find that in order for the above equation to hold we must require that
\be\label{cutoff2}
W_{\t;ax}\dot\R^\t_{xy}W_{\t;yb}=-W_{\t_{ab};ab}\d(\t-\t_{ab})\ .
\ee
The above equation will allow us to integrate the NPRG flow. In Section \ref{sec:causalityII} we analyze in detail the causal structure of the resulting flow equations and their solutions. In the end we show that the resulting solutions for $\Gamma^{(n)}_\t$ are consistent with eq.~(\ref{Gamma_causal}), and therefore causality is preserved.

\subsection{Solving the NPRG flow equations. The Helmholtz hierarchy.}\label{sec:Solving}
Following the results from Section \ref{sec:IC} the initial conditions for the NPRG flow with a temporal cutoff are given by 
\be\label{ICt}
\Gamma^{(n)}_{\eta_I}[\vf_{\t}]=S^{(n)}[\vf_{\t}]\ .
\ee
Note that above we restored the cutoff index for $\vf$. From (\ref{S}) and (\ref{ICt}) one can see that the initial conditions for the flow are such that the only non-zero functional derivatives $S^{(n)}$ are for $n\leq 3$.

From (\ref{S}) we find
\be\label{IC1}
&&\Gamma_{\eta_I;\bar f_{\t,a}}=\bar \chi_{\t,x}\L{x}{a}-2\bar \chi_{\t,x} K_{xya}\bar f_{\t,y}\nonumber\\
&&\Gamma_{\eta_I;\bar\chi_{\t,a}}=\L{a}{x}\bar  f_{\t,x}-K_{axy}\bar f_{\t,x}\bar f_{\t,y}-\mu_a-\Delta_{ax}\bar \chi_{\t,x}\ .
\ee
The initial condition, $\Gamma^{(2)}_{\t=\eta_I}$, obtained from (\ref{S}) is:
\be\label{IC2}
\renewcommand{\arraystretch}{1.8}
\Gamma_{\eta_I;ab}=\left(
\begin{array}{cc}
-2\bar\chi_{\t,x} K_{xab} & \L{b}{a} -2 K_{bxa}\bar f_{\t,x} \\
 \L{a}{b}-2 K_{axb}\bar f_{\t,x} & -\Delta_{ab}
\end{array}
\right).
\ee
Later we show (cf. eq.~(\ref{chi0})) that $\bar \chi_{\t,x}$ can be consistently set to zero.

Finally, the initial condition, $\Gamma^{(3)}_{\t=\eta_I}$, obtained from (\ref{S}) is $\Gamma_{\eta_I;abc}=0$ unless one of the $\phi_i$ ($i=a,b,c$) fields is $\chi$, and the other two are $f$. In that case $\Gamma_{\eta_I;abc}=-2K_{\pi(abc)}$,
where $\pi(abc)$ is the permutation of $(abc)$ that puts the $\bar\chi$ field index first. For example, $\Gamma_{\eta_I;\bar f_a\bar\chi_b\bar f_c}=-2K_{\pi(abc)}=-2K_{bac}$.

The generation of any non-trivial Fokker-Planck-like terms in the equation for the average field, $\la \phi\ra$, is captured by the functional flow at second order in the functional expansion. Moreover, any nontrivial evolution of the bare vertex $K_{abc}$ is captured at third order. Thus, the truncation we impose on the NPRG flow is
\be\label{trunc}
\Gamma^{(n\geq4)}_{\t}=\Gamma^{(n\geq4)}_{\eta_I}=0\ .
\ee
The truncation above is simply for convenience. The method can be easily extended to include nonzero $\Gamma^{(n\geq4)}_{\t}$. As $\Gamma^{(n\geq4)}_{\t}$ involves at least $n$ bare vertices\footnote{This can be seen by considering the number of $\Gamma^{(3)}$'s entering in the $n$-th order of the NPRG flow.}, $K$, this means that 
\be\label{GammanOrder}
\Gamma^{(n\geq 4)}_{\t}\sim \mathcal{O}(\varepsilon^n)\ ,
\ee
including higher order terms, where $\varepsilon$ is the HH ordering parameter, (\ref{ordering}). Thus, (\ref{trunc}) corresponds to a truncation at $\mathcal{O}(\varepsilon^4)$.

Combining (\ref{NPRG}), (\ref{Wt}), (\ref{cutoff2}), (\ref{ICt}) and (\ref{trunc}) it is straightforward to integrate the NPRG flow. 
We only need to specify a boundary condition for $\Gamma_{\tau;a}$ at $\tau_a$. The cutoff function vanishes at $\tau_a$, and from (\ref{j}), setting the external sources to zero (as per the discussion in Section \ref{subs:eff_act}), we obtain
\be\label{EqOfM}
\Gamma_{\t_a;a}[\vf_{\t_a,a}]=0\ ,
\ee
which is the equation of motion for the average field, $\vf$.
Thus, from the first order NPRG equation, combined with (\ref{EqOfM}) we obtain the first equation of the Helmholtz hierarchy:
\be\label{hie1}
\Gamma_{\eta_I;a}[\vf_{\t_a,a}]=-\frac{1}{2}\Gamma_{\t_{12};a12}W_{;21}\theta^1(\t_a-\t_{12})\ ,
\ee
where we used that $W_{\t_{ab};ab}=W_{;ab}$. We denoted the dummy integration subscripts above with Arabic numerals, which will make the subscripts below easier to follow. Combining equations (\ref{IC1}) and (\ref{hie1}) we obtain an equation for $\vf_{\t,a}$ which is necessary for obtaining $\Gamma_{\eta_I;ab}$. We show later (cf. eq.~(\ref{chi0})) that in order to be consistent with causality we need $\bar\chi_{\t,a} =0$. 

For $\Gamma^{(2)}_\t$ we obtain
\be\label{hie2a}
\Gamma_{\t_{ab};ab}-\Gamma_{\eta_I;ab}=\left.-\frac{1}{2}\Gamma_{\tilde\t;a12}\Gamma_{\tilde\t;b34}W_{;23}W_{;41}\theta^1(\t_{ab}-\tilde\t)\right|_{\tilde\t=\t_{1234}}\ .
\ee
We will prove that the term $\mathcal{O}(\Gamma^{(4)})$ in the above equation vanishes identically due to causality and the structure of the bare action. And in general, for $n>1$ we will find that $\Gamma^{(n)}$ depends only on the $\Gamma$'s up to $\Gamma^{(n+1)}$, and not on $\Gamma^{(n+2)}$. We show this in Appendix \ref{app:zeroblob}.

Finally, for $\Gamma^{(3)}_\t$, the flow equation can be integrated to give the third Helmholtz equation
\begin{flalign}\label{hie3}
\Gamma_{\t;abc}- & \Gamma_{\eta_I;abc}=\\
\nonumber
& =\left.\frac{1}{2}\Gamma_{\tilde\t;a12}\Gamma_{\tilde\t;b34}\Gamma_{\tilde\t;c56}W_{;23}W_{;45}W_{;61}\theta^1(\t-\tilde\t) \ + \  \left(b\leftrightarrow c\right)\right|_{\tilde\t=\t_{123456}}+\mathcal{O}(\Gamma^{(4)})\ ,
\end{flalign}
with $\eta_I<\t\leq\t_{abc}$, the last inequality arising from eq.~(\ref{Gamma_causal}). Again, we used that the term $\mathcal{O}(\Gamma^{(5)})$ vanishes identically as per Appendix \ref{app:zeroblob}. The causal structure of the Helmholtz hierarchy is discussed in detail in Section \ref{sec:causalityII}. There we show that the above equations are consistent with (\ref{Gamma_causal}), which is the reason why we integrated the flow up to the maximum time argument.

The truncated hierarchy above is not closed unless we specify $W_{;ab}$. Combining (\ref{Winv}), 
(\ref{cutoff}) and (\ref{Wt}) for time arguments larger than $\eta_I$ we obtain
\be\label{inversefull}
W_{;ab}=W_{\t_{ab};ab}&=&W_{\t_{ab};ax}\Gamma_{\t_{ab};xy}W_{\t_{ab};yb}+W_{\t_{ab};ax}\R^{\t_{ab}}_{xy}W_{\t_{ab};yb}=W_{\t_{ab};ax}\Gamma_{\t_{ab};xy}W_{\t_{ab};yb}\nonumber\\
&=&W_{\t_{ax};ax}\Gamma_{\t_{xy};xy}W_{\t_{yb};yb}\theta^1(\t_{ab}-\t_{xy})\ .
\ee
To obtain the third equality above we used that $W_{\t_{ab},xy}\propto \theta^1(\t_{ab}-\t_{xy})$ and $\R^{\t_{ab}}_{xy}\propto \theta^0(\eta_x-\t_{ab})\d(\eta_x-\eta_y)$, paying attention to the superscripts of the $\theta$-functions. For the last equality above we used (\ref{Gamma_causal}) and (\ref{Wt}).
Using (\ref{Wt}) and (\ref{inversefull}), the second equation, eq.~(\ref{hie2a}), in the hierarchy can be written as
\be\label{hie2}
W_{;ab}&=&W_{\t_{ab};ab}=W_{;ax}\Gamma_{\eta_I;xy}W_{;yb}\theta^1(\t_{ab}-\t_{xy})-\\
&-&\left. \frac{1}{2}W_{;ax}\Gamma_{\tilde\t;x12}W_{;23}W_{;41}\Gamma_{\tilde\t;y34}W_{;yb}\, \theta^1(\t_{ab}-\t_{xy}) \theta^1(\t_{xy}-\tilde\t)\right|_{\tilde\t=\t_{1234}}\nonumber\ .
\ee
Note that the structure of the above equations is the same as that of the flow equations of \cite{Gasenzer:2007}, where the sharp temporal cutoff was first proposed. 

Equations (\ref{hie1}), (\ref{hie2a}), (\ref{hie3}) and (\ref{hie2}) are the final result of this subsection. The first three of them represent the first three equations of the HH. Showing that the solution of (\ref{hie2}) for $W_{\t_{ab};ab}$ depends only on the statistical properties of the system at previous times 
will be done after we write down the flow equations in terms of Feynman diagrams.

\subsection{Diagrammatic approach}

In this section we introduce a diagrammatic approach for writing down the NPRG and Helmholtz hierarchies. It will allow us to simplify the flow equations; to easily check all causality constraints; and to show that the auxiliary field remains unobservable after obtaining the truncated Helmholtz hierarchy with the choice of a cutoff made in sec.~\ref{sec:causalityI}.

The NPRG flow equations can be easily cast in Feynman diagrams. Indices corresponding to $\chi$ we write as wiggly lines; while those corresponding to $f$ -- with straight lines. The functional derivatives of $W$ we denote with black lines, while those of $\Gamma$ -- with gray lines and gray dots/blobs. As an example, consider $W^{(2)}$ which can be written as
\vspace{0.5cm}
{\unitlength=1mm
\begin{fmffile}{d001}
\be
W_{\t;ab}\equiv\left(
\begin{array}{cc}
\parbox{26mm}{\fmfframe(2,0)(4,0){\begin{fmfgraph*}(20,15)
\fmfleft{i} \fmfright{o}
\fmf{plain,label=$G_{\t;ab}$,l.side=left}{i,o}
\fmflabel{$b$}{o}
\fmflabel{$a$}{i}
\end{fmfgraph*}}}  
 & 
\parbox{26mm}{\fmfframe(4,0)(2,0){\begin{fmfgraph*}(20,15)
     \fmfleft{i} \fmfright{o}
\fmflabel{$b$}{i} \fmflabel{$a$}{o}
     \fmf{phantom,label=$\RRl{\t;a}{b}$,tag=1,l.side=left}{i,o}
\fmffreeze
\fmfipath{p[]}
\mypath{1}{i}{o}
   \end{fmfgraph*}}}
 \\
\parbox{26mm}{\fmfframe(2,0)(4,0){\begin{fmfgraph*}(20,15)
     \fmfleft{i} \fmfright{o}
\fmflabel{$a$}{i} \fmflabel{$b$}{o}
     \fmf{phantom,label=$\RR{\t;a}{b}$,tag=1,l.side=left}{i,o}
\fmffreeze
\fmfipath{p[]}
\mypath{1}{i}{o}
   \end{fmfgraph*}}}
  & 
\parbox{26mm}{\fmfframe(4,0)(2,0){\begin{fmfgraph*}(20,15)
     \fmfleft{i} \fmfright{o}
\fmflabel{$a$}{i}\fmflabel{$b$}{o}
     \fmf{wiggly}{o,i}
   \end{fmfgraph*}}}
\end{array}
\right)\ .
\ee
\end{fmffile}}\noindent
{\flushleft The} above equation combined with (\ref{Wtable}) defines the linear response function, $\RR{\t;a}{b}$, and the two-point correlation function $G_{\t;ab}$. The arrow above $\RR{\t;a}{b}$ indicates the direction in time of propagation of the linear response (i.e. $\eta_b> \eta_a$ in this case); and thus we define $\RRl{\t;b}{a}\equiv \RR{\t;a}{b}$.

To give an example for $\Gamma^{(3)}$, the vertex $\Gamma_{\t;\chi_af_bf_c}$ is represented by:
\vspace{0.5cm}
{\unitlength=1mm
\begin{fmffile}{d002}
\be
\Gamma_{\t;\bar \chi_a\bar f_b\bar f_c}&\equiv &
\parbox{34mm}{\fmfframe(10,5)(4,5){\begin{fmfgraph*}(20,20)
\fmfleft{i} \fmfrightn{o}{2}
\fmf{wiggly,foreground=0.3*black+0.7*white}{i,v}
\fmf{plain,foreground=0.3*black+0.7*white}{o1,v}
\fmf{plain,foreground=0.3*black+0.7*white}{o2,v}
\fmflabel{$a$}{i}
\fmflabel{$b$}{o1}
\fmflabel{$c$}{o2}
\fmfv{decor.shape=circle,decor.filled=30,decor.size=3thick,foreground=0.3*black+0.7*white}{v}
 \end{fmfgraph*}}}
\ee
\end{fmffile}
}\noindent
{\flushleft Note} that $\Gamma^{(n)}_\t$ and $W^{(n)}_\t$ are symmetric, since the functional derivatives commute, e.g. $\Gamma_{\t;\bar \chi_a\bar f_b\bar f_c}=\Gamma_{\t;\bar f_b\bar \chi_a\bar f_c}$.

\subsection{Causality of the flow II}\label{sec:causalityII}

Now let us show explicitly using the diagrammatic approach introduced in the last section that the truncated NPRG/Helmholtz hierarchy is causal and that the auxiliary field is unobservable.

To show that, first we prove several properties of $\Gamma^{(n)}_\t$. Let us for now assume that all 1PI vertices whose external legs are all straight lines (corresponding to functional derivatives in $\bar f$) are zero:
\be\label{assumed_for_now}
\Gamma_{\t;\bar f_1\bar f_2\cdots \bar f_n}=0\ .
\ee
We will prove this relation shortly. Using (\ref{inversefull}), we see that since $\Gamma_{\t;\bar f_1\bar f_2}=0$, we obtain that the wiggly propagator vanishes:
\be\label{wiggly_prop}
W_{\t;\bar \chi_a\bar \chi_b}=0\ ,
\ee
where for brevity we used subscripts $\bar \chi$ instead of $j_{\bar \chi}$. We will use the same short-hand notation for $W$ from now on.
Then, we can show that the following relation holds:
\newline
\begin{myfigure}
\includegraphics[scale=1.125]{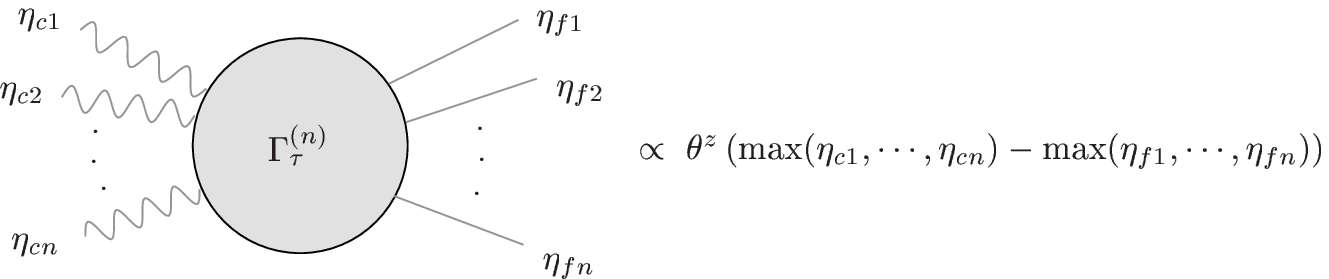}
\caption{\label{n_vert_Gamma}}
\end{myfigure}
\vspace{-0.5cm}{\flushleft where} $z=0$ if the corresponding bare vertex vanishes, i.e. $\Gamma^{(n)}_{\eta_I}=0$; and $z=1$ otherwise.
Let us check that the above relation is consistent with the Helmholtz hierarchy. From (\ref{assumed_for_now}) we know that there must be at least one wiggly external leg to any 1PI diagram. Thus, we can see that the 1PI $m$-point function with at least one wiggly and one straight external leg is given by
\vspace{0.5cm}
\begin{myfigure}
\includegraphics[scale=1.165]{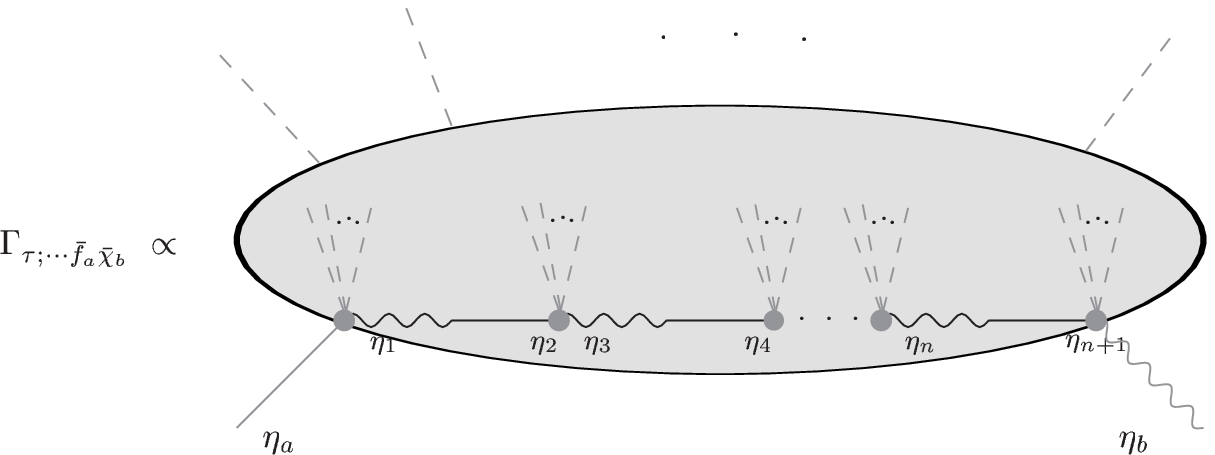}
\caption{\vspace{1.05cm}\label{series_vert_Gamma}}
\end{myfigure}
{\flushleft where} a dashed line stands for either a wiggly or a straight line. The blob represents the 1PI vertex, $\Gamma^{(m)}_\t$. In the above equation we used (\ref{wiggly_prop}) which tells us that to any wiggly external leg of a 1PI diagram one can attach only the response function $\RRl{\t,a}{b}$. The dashed lines drawn inside the large blob can be attached to either internal propagators, or can be external legs. 

The causality condition (\ref{CC}) combined with (\ref{response}) implies 
\be\label{Rab_cause}
\RRl{\t;a}{b}\propto\theta^0(\eta_a-\eta_b)\ ,
\ee
which will be checked to be consistent with the NPRG flow further below.
According to (\ref{n_vert_Gamma}), without loss of generality, we can choose each wiggly leg shown inside the blob in (\ref{series_vert_Gamma}) to correspond to the maximum time of the corresponding vertex. For the rightmost vertex in the blob we have two choices: the external leg of latest time of that vertex can either attach to an internal line in the blob, or can correspond to the external leg with time $\eta_b$. In the former case, the chain of $\RRl{\t;a}{b}$ propagators must eventually close on itself, thus forming a closed internal time-loop, which must vanish as per (\ref{Rab_cause}). In the latter case, using (\ref{Rab_cause}) we can write:
\be
\eta_b\geq \eta_{n+1}> \eta_n \geq \cdots \geq \eta_4>\eta_3 \geq \eta_2> \eta_1 \geq \eta_a\ .
\ee
Thus, the NPRG flow does not violate (\ref{n_vert_Gamma}) if it holds for some $\tau$. Equation (\ref{n_vert_Gamma}) holds for $\tau=\eta_i$ since the bare action includes only equal-time operators. Therefore, by induction, we conclude that (\ref{n_vert_Gamma}) holds for all $\tau$.

We are now in a position to prove eq.~(\ref{assumed_for_now}). The proof goes in complete analogy with the above proof of eq.~(\ref{n_vert_Gamma}). We can write the vertex $\Gamma_{\t;\bar f_1\bar f_2\cdots \bar f_n}$ as in (\ref{series_vert_Gamma}), except $\eta_b$ must correspond to a straight external line. Since all external lines must be straight, the chain of $\RRl{\t;a}{b}$ propagators inside the blob must eventually close on itself, which means that the diagram vanishes\footnote{For this to be the case, note that it is crucial that $\theta(0)=0$ in (\ref{Rab_cause})}, if it vanishes at some $\t$. Since eq.~(\ref{assumed_for_now}) holds at $\tau=\eta_i$ if $\bar \chi_\t=0$, by induction it holds for all $\t$.

So, now we have to see whether $\bar \chi_\t$ vanishes. Its equation of motion is given by the $\bar f$ component of eq.~(\ref{hie1}). According to Appendix \ref{app:zeroblob}, the 3-point 1PI vertex entering in (\ref{hie1}) reduces to the bare vertex. Since the bare vertex contains only equal-time operators, from (\ref{assumed_for_now},\ref{wiggly_prop},\ref{n_vert_Gamma},\ref{Rab_cause}) we see that the right hand side of (\ref{hie1}) must vanish, and therefore 
\be\label{chi0}
\bar \chi_{\t}=0\ ,
\ee
as required by the discussion above.

Next we want to prove that the truncated NPRG hierarchy indeed satisfies the causality condition, eq.~(\ref{CC}), which can be rewritten using (\ref{response},\ref{Wcum}) as:
\be\label{W_cause}
W_{\t;\bar f_{b_1},\cdots,\bar f_{b_m}\bar \chi_{a_1},\cdots,\bar \chi_{a_n}}\propto\, \theta^0\big[\max(\eta_{b_1},\cdots,\eta_{b_m})-\max(\eta_{a_1},\cdots,\eta_{a_n})\big]\ .
\ee
To obtain $W^{(n)}_\t$ from $\Gamma^{(m)}_\t$ ($m\leq n$), one has to take functional derivatives of (\ref{inverseW2}) with respect to $j$ at fixed $\tau$. As an example, for $n=3$ we obtain
\be\label{3vert}
W_{\t;abc}=-W_{\t;ax}W_{\t;by}W_{\t;cz}\Gamma_{\t;xyz}\ .
\ee
Since $W$ generates all connected diagrams, while $\Gamma$ generates all 1PI diagrams, we can see that in order to obtain a cumulant or a response function, $W^{(n)}_\t$, we must chain one or more $\Gamma^{(m)}_\t$ with internal propagators and attach propagators, $W^{(2)}_\t$, to the external legs of the resulting diagram. Therefore, if we want to obtain one of the response functions, $W_{\t;\bar \chi_{a_1}\cdots\bar \chi_{a_l} \bar f_{b_1}\cdots \bar f_{b_m}}$, from (\ref{wiggly_prop}) we can see that each of the external $\eta_a$ (corresponding to the derivatives of $W$ in $\bar \chi_{a_s}$) must belong to a propagator $\RRl{\t;w}{a}$ with $\eta_a<\eta_w$, where $w$ attaches to the straight external leg of a 1PI vertex. According to (\ref{assumed_for_now}), that vertex has at least one external wiggly leg with time $\eta_x$, such that $\eta_x\geq \eta_w$. That leg must attach to $\RRl{\t;y}{x}$ with $\eta_y>\eta_x$, which is either external or in turn attaches to another 1PI vertex, thus forming a chain of causally ordered 1PI vertices. In the end, this causally ordered sequence implies that $\eta_{b_r}>\eta_{a_s}$ for some $r=1, \dots, m$ and all $s=1, \dots, l$. Thus, we showed that eq.~(\ref{W_cause}), and hence  eq.~(\ref{Rab_cause}), is consistent with the NPRG flow, and by induction it must hold. Thus, we showed that the NPRG flow is causal.

Next we must show that the field $\chi_\t$ is unobservable. We already showed that its mean value is zero, eq.~(\ref{chi0}). Now we have to show that all of its cumulants vanish. Combining (\ref{ghost}) and (\ref{Wcum}), that is equivalent to showing that:
\be\label{W_unobs}
W_{\t;\bar \chi_{1},\cdots,\bar \chi_{n}}=0\ .
\ee
Performing the same analysis as we did above to show the causality of the flow, we can see that all $\chi$ cumulants must contain a 1PI diagram with external legs which are all straight. That diagram vanishes, (\ref{assumed_for_now}), and therefore we conclude that indeed eq.(\ref{W_unobs}) holds.

The whole discussion above is self-consistent and consistent with the NPRG flow if both the bare action contains equal time operators, and eq.~(\ref{Gamma_causal}) holds. The former condition holds for the CDM. The latter condition leads to eq.~(\ref{cutoff2}), which introduces the delta function in the NPRG flow equations, which allowed us to integrate the flow. 

So, we are left to prove eq.~(\ref{Gamma_causal}). If we assume it is true, then we know that the delta function in eq.~(\ref{cutoff2}) tells us that 
\be\label{schematicGn}
\Gamma^{(n)}_\t\propto \theta^{1}(\t-\tilde \t)\ ,
\ee
where $\tilde \t$ stands for the maximum of the time parameters of all internal line propagators. The proportionality factor above does not depend on $\t$. Thus, the only way that $\t$ appears on the right hand side above is as an upper bound on the integrals over the time parameters of the internal lines. Let us denote the maximum of the time parameters of the external legs of $\Gamma^{(n)}_\t$ as $\eta_{\mathrm{max}}$. If we can prove that $\eta_{\mathrm{max}}\geq \tilde \t$, then for $\t\geq \eta_{\mathrm{max}}$, the right hand side of eq.~(\ref{schematicGn}) will remain constant, thus proving eq.~(\ref{Gamma_causal}). 

So, let us prove that 
\be
\eta_{\mathrm{max}}\geq \tilde \t\ .
\ee
 Let us assume to the contrary: there exists an internal propagator with a time parameter which succeeds all external leg time parameters. That time parameter can be on either an internal wiggly line or on an internal straight line. If it is on a straight line, then according to (\ref{assumed_for_now}) and (\ref{n_vert_Gamma}), there must exist a wiggly line in the diagram which comes later. It cannot be an external leg, or otherwise our assumption will be violated. So, it must be that the latest internal time parameter must be on an internal wiggly line. However, that wiggly line can attach only to the internal propagator $\RRl{\t;a}{b}$ due to eq.~(\ref{wiggly_prop}). And therefore, we see that there exists a straight internal line which has the latest internal time parameter. But we already saw that this is impossible. Thus, our assumption must be wrong, and therefore $\eta_{\mathrm{max}}\geq \tilde \t$. Therefore, eq.~(\ref{Gamma_causal}) holds.\footnote{Eq.~(\ref{Gamma_causal}) was proved in \cite{gazpaw2} using similar arguments.}

One last property of $\Gamma^{(n)}_\t$ deserves our attention. For $\t<\eta_{\mathrm{max}}$ we can prove using an analysis similar to the above (see Appendix B of \cite{gazpaw2}) that
\be\label{Gamma_tau_small}
\Gamma^{(n)}_\t=\Gamma^{(n)}_{\eta_I}\ , \ \ \hbox{for }\t<\eta_{\mathrm{max}}\ .
\ee
Thus, the flow of $\Gamma^{(n)}_\tau$ turns out to be rather trivial under the sharp temporal cutoff: $\Gamma^{(n)}_\tau=\Gamma^{(n)}_{\eta_I}=S^{(n)}$ for $\t<\eta_{\mathrm{max}}$, and jumps to $\Gamma^{(n)}_\tau=\Gamma^{(n)}$ for  $\t\geq \eta_{\mathrm{max}}$.

Now let us show that $W^{(n)}_\t$ depends on the statistical properties of the system only at previous times. To see this, note that each 1PI vertex entering in the cumulants, $W^{(n)}_\t$, must have an external wiggly line, which corresponds to the latest time parameter (as shown above) entering in that 1PI diagram. We must chain at least one 1PI diagrams with internal propagators, satisfying (\ref{wiggly_prop}). Thus, we end up with an external wiggly line which has the largest time parameter, $\eta_x$, among the time parameters in the chain. The only external propagator that can attach to that line is $\RRl{\t;a}{x}$, which has $\eta_a>\eta_x$, with $\eta_a$ being the time of the external leg of $W^{(n)}_{\t;\cdots \bar f_a}$. Thus, $W^{(n)}_\t$ depends only on the past properties of the system, which can be seen in e.g. eq.~(\ref{inversefull}).

This concludes our proof that under the sharp temporal cutoff, the structure of the NPRG flow and its solutions, the Helmholtz hierarchy, is consistent with causality; and that the auxiliary field, $\chi$, is unobservable. This property is preserved under truncation of the hierarchy as long as any truncation to the NPRG (or Helmholtz) hierarchy is consistent with all causality conditions discussed above.

\subsection{Summary and discussion}\label{sec:summary_disc_6}

In this section we integrated the NPRG flow equations for the CDM using the sharp temporal cutoff, proposed in \cite{Gasenzer:2007} for dealing with non-equilibrium quantum field dynamics. The solution is what we call the Helmholtz hierarchy, the first three orders of which are given by equations (\ref{hie1},\ref{hie2a},\ref{hie3}) supplemented by (\ref{IC1},\ref{IC2},\ref{hie2},\ref{chi0}).

We showed that the presence of the cutoff $\tau$ in $\Gamma^{(n)}_\t$, implies that the time parameters in all internal loops are bounded from above by $\t$. If $\t=\eta_I$, we directly obtain the anticipated result: $\Gamma^{(n)}_{\t=\eta_I}=S^{(n)}$. Thus, by moving $\t$ we interpolate between the bare action $S$ for the CDM and the fully integrated effective action, $\Gamma$, by including gravitational interactions up until the cutoff in time, $\tau$. The flow of $\Gamma^{(n)}_\tau$ turns out to be a simple jump from $S^{(n)}$ to $\Gamma^{(n)}$ at $\t=\eta_{\mathrm{max}}$ (the largest time parameter of the external legs of $\Gamma^{(n)}$).

Truncating the HH at $\mathcal{O}(\Gamma^{(n)})$ for $n\geq 4$ is in general a better approximation than truncating it at $\mathcal{O}(\varepsilon^n)$ since $\Gamma^{(n)}$ is a sum of terms which are $\OK{n}\sim\mathcal{O}(\varepsilon^n)$ or higher. However, a simple $\Gamma^{(n)}$ truncation is not physical, as there is no manifest physical ordering parameter. So, the way one should approach the HH is by first truncating at $\mathcal{O}(\Gamma^{(n)})$ for $n\geq 4$, and then truncate at $\mathcal{O}(\varepsilon^n)$. For $n<4$, one should be careful with the bare values contributing to $\Gamma^{(n)}$ (e.g. at lowest order $\Gamma_{;\bar \chi \bar f\bar f}= \Gamma_{\eta_I;\bar \chi \bar f \bar f}\sim \mathcal{O}(\varepsilon)$).

We showed explicitly that the truncated NPRG and Helmholtz hierarchies obey causality, equations (\ref{CC}) and (\ref{Wt}), to all orders. Moreover, we showed that the generated auxiliary field is unobservable, as required by eq.~(\ref{ghost}). 

Combining eq.s~(\ref{IC1},\ref{hie1},\ref{hie3}) we find the equation of motion for $\bar f$:
\be\label{f1}
\L{a}{x}\bar f_x-K_{axy}\bar f_x \bar f_y-\mu_a=K_{axy}G_{xy}\ ,
\ee
where we used that the right hand side of eq.~(\ref{hie3}) vanishes exactly as per eq.~(\ref{blobequation}) in Appendix \ref{app:zeroblob}.
Thus, the above equation is exact, and note that it is identical to the first extended BBGKY equation, eq.~(\ref{proba2}).

The two-point correlation function can also be solved for easily to obtain (using (\ref{Wtable}) and (\ref{inversefull})):
\begin{myfigure}
\includegraphics[scale=1]{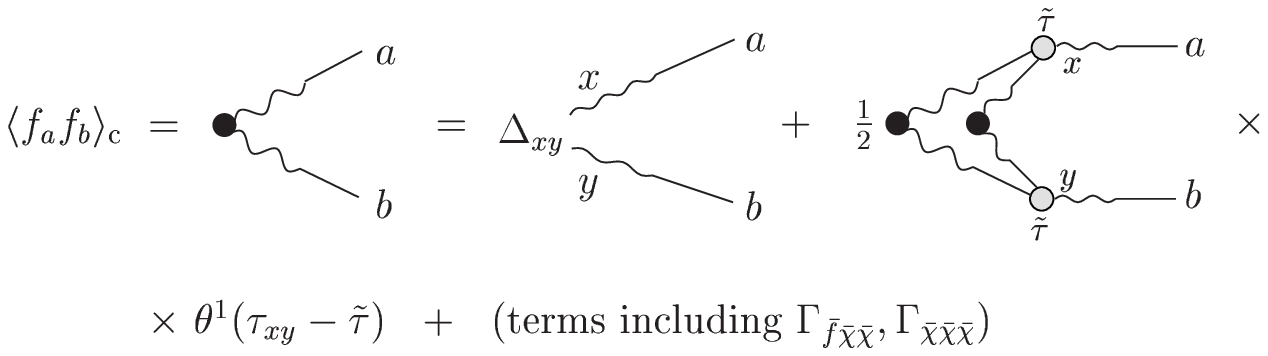}
\caption{\vspace{-0.1cm}\label{2pt_corr}}
\end{myfigure}
\noindent where the first equality defines the black blob as the amputated two-point correlation function. The diagrams above are drawn so that time flows along the horizontal axis in each diagram. That is done to highlight the causal structure of the diagrams. Thus we see that the two-point function is generated by two distinct contributions: The first term in (\ref{2pt_corr}) corresponds to the initial conditions propagated forward in time; while the rest of the terms show the generation of power through mode coupling. A result with the same structure was obtained in Renormalized Perturbation Theory (RPT) (e.g. eq.~(9) in \cite{Crocce:2007dt}). As an example of how mode coupling generates power in the two-point function, in the second term above we can see that two pairs of modes are generated (at the black blobs) either by propagating the initial conditions or by nonlinearities, and then those pairs of modes interact gravitationally (at the gray vertices) to dump power to the two point function.

The above equation can be simplified using the $\theta$ function in eq.~(\ref{2pt_corr}) as follows:
\be\label{2ptFun}
G_{ab}&=&\RRl{a}{x}\Delta_{xy} \RR{y}{b}+\frac{1}{2}\RRl{a}{x}\Pi_{xy}\RR{y}{b}\\
\Pi_{xy}&\equiv&4 K_{x12}G_{13}G_{24}\left(-\frac{1}{2}\Gamma_{\t_{12};\bar f_3\bar f_4\bar\chi_y}\right)\theta(\t_{12}-\t_{34})+(x\leftrightarrow y)+\nonumber\\
&&+(\hbox{terms including }\Gamma_{;\bar f\bar \chi\bar \chi}\ \hbox{and} \ \Gamma_{;\bar \chi\bar \chi\bar \chi})\label{pi1}\ .
\ee 
As in (\ref{2pt_corr}) the term $\overleftarrow R_{ax} \Delta_{xy} \overrightarrow R_{yb}$ tells us how the initial 2-point function is propagated forward in time by the response function, $\overleftarrow R$. The term $\overleftarrow R_{ax} \Pi_{xy} \overrightarrow R_{yb}$ tells us how much power is generated from mode coupling due to the nonlinear evolution of CDM. Thus, $\Pi$ is usually referred to as the ``self-energy'' of the system.  Note that not surprisingly eq.~(\ref{2ptFun}) has an identical structure to the equation for the two-point function in RPT (the second equation in Fig.~10 of \cite{Crocce:2005xy}).

Truncating the above equation at $\mathcal{O}(\varepsilon^3)$, the first brackets in (\ref{pi1}) become $K_{y34}$, and the last line of that equation drops out. Thus, one obtains
\be\label{pi}
\Pi_{xy}=4 K_{x12}G_{13}G_{24}K_{y34}\ .
\ee
Equations (\ref{2ptFun}) and (\ref{pi}) are equivalent to eq.~45 of \cite{Valageas:2003gm} (after using his eq.~47; keeping the contribution of the initial conditions; and changing to his set of variables). One important difference from \cite{Valageas:2003gm}, however, is that to be consistent, we have to truncate (\ref{2ptFun}) and (\ref{pi}) at $\mathcal{O}(\varepsilon^3)$.

The linear response function (referred to as the ``nonlinear propagator'' in \cite{Crocce:2005xy}) is given by
\begin{myfigure}
\includegraphics[scale=1.18]{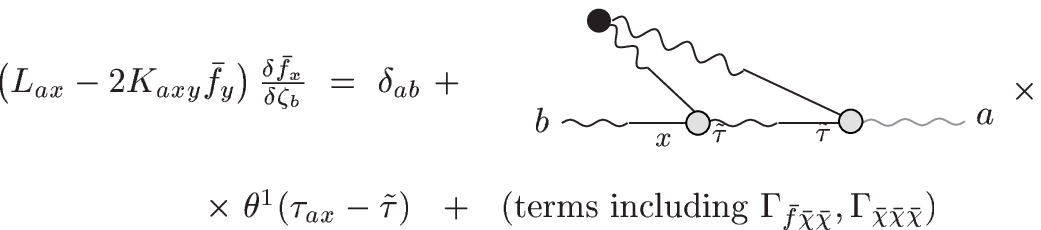}
\caption{\vspace{-0.15cm}\label{2pt_resp}}
\end{myfigure}
\noindent Note that the label $a$ in the diagram above lies on a gray wiggly line. The $K$ term on the left hand side shows us how the disturbance generating the linear response couples to the average CDM distribution, $\bar f$. The term represented by the diagram, shows how the response function is modified by mode coupling, where a pair of modes (generated at the black blob) interact with the response function at the vertices (gray vertices). The structure of the above equation is the same as that in RPT (the first equation in Fig.~10 of \cite{Crocce:2005xy}).
Using the causal structure of the vertices in analogy as we did in Appendix \ref{app:zeroblob}, the above equation combined with the equation for the other non-diagonal element of $W^{(2)}$ (see (\ref{Wtable})) can be written as
\be
&&\left(\L{a}{x}-2K_{axy}\bar f_{y}\right)\RRl{x}{b}=\delta_{ab}+\overleftarrow{\Sigma_{ax}}\RRl{x}{b}\label{responseFun}\\
&&\left(\L{x}{a}-2K_{xya}\bar f_{y}\right)\RR{x}{b}=\delta_{ab}+\overrightarrow{\Sigma_{ax}}\RR{x}{b}\label{responseFun1}\\
\overrightarrow{\Sigma_{ab}}&\equiv&4\left(-\frac{1}{2}\right)\Gamma_{\t_{34};\bar f_a\bar\chi_1\bar f_2}\RR{1}{3}G_{24}K_{b34}+ (\hbox{terms including}\Gamma_{;\bar f\bar \chi\bar \chi}\ \hbox{and} \ \Gamma_{;\bar \chi\bar \chi\bar \chi})\ .\nonumber\\
\label{SIGMA}
\ee
The term containing $\overrightarrow{\Sigma}$ -- the ``damping self-energy'' -- serves a similar role as $\Pi$, i.e. it redistributes power due to the nonlinear interactions. Note that the above expression for the damping self-energy is equivalent to the expression obtained by \cite{Valageas:2003gm} (cf. his eq. 56).

At $\mathcal{O}(\varepsilon^3)$, the above set of equations is identical to the result in \cite{Valageas:2003gm} (his eq.~(46),(47)), reducing the damping self-energy to:
\be\label{sigma}
\overrightarrow{\Sigma_{ab}}&=&4K_{12a}\RR{1}{3}G_{24}K_{b34}\ .
\ee
Again note that to be consistent with our physical expansion,  we have to truncate (\ref{responseFun}), (\ref{responseFun1}) and (\ref{sigma}) at $\mathcal{O}(\varepsilon^3)$, unlike \cite{Valageas:2003gm}.


As a check of our results, let us see how one can recover the final result by Valageas \cite{Valageas:2003gm} from the HH. If one does not truncate our equations with respect to $\varepsilon$, but instead work to $\mathcal{O}(P_L^2)$, one recovers the result (eq.~(42)-(50)) obtained by \cite{Valageas:2003gm} using the steepest-descent method applied to a large-$N$ expansion of the action. Thus, if instead of using $\varepsilon$, we use $P_L$ as our ordering parameter, we can recover the result of second order standard perturbation theory \cite{jain}, which result was re-derived by Valageas \cite{Valageas:2003gm} using his equations~(42)-(50) and initial conditions from which important phase-space information has been dropped (see discussion in Section \ref{sec:zeldovich_IC}) by the $P_L$ expansion.

Now let us see how the HH recovers information about stream crossing. The easiest way to show that is to work to zero order in $\varepsilon$, which is equivalent to the ZA. In that case we have
\be
G_{ab}&=&\RRl{a}{x}\Delta_{xy} \RR{y}{b}\label{G_Zel}\\
\RRl{a}{x}&=&\theta(\eta_a-\eta_b) \delta_D(\bm{k}_a-\bm{k}_b)\delta_D\bigg(\bm{w}_a-\bm{w}_b+\big(D(\eta_a)-D(\eta_b)\big)\bm{k}_a\bigg)\label{R_Zel}\ .
\ee
Note that the last equation is equivalent to (\ref{Rabz}) as must be the case. The only difference between the exact two-point function (\ref{2pfz}) in the ZA and $\langle f_af_b\rangle_{\mathrm{c}}$ entering in $\Delta_{ab}$ is the time arguments. That difference is removed by the Dirac delta functions in $\RRl{a}{x}$ in the expression for $G_{ab}$ above. Thus, $G_{ab}$ above readily reduces to (\ref{2pfz}). Therefore, we can see that the HH at zero order in $\varepsilon$ directly recovers the exact 2-point function in ZA. Thus, we can conclude that the HH preserves the information about stream crossing. 

From (\ref{G_Zel},\ref{R_Zel}) one can also see that the high $k$ decay kernel in the two-point function (\ref{2pfz_simple}) is entirely due to the presence of the decay kernel in the initial conditions, $\Delta_{ab}$, where the decay in $\o$ is translated to a decay in $\k$ by the delta functions in $\RRl{a}{x}$. The structure of $\RRl{a}{x}$ is such that the decay at high $\o$ of the zero order $\bar f$, (\ref{fbarZ}), and $\Delta_{ab}$ will be converted to a decay at high $k$ of the CDM power spectrum to all orders in $\varepsilon$. Thus, the behavior of the \textit{density} response function at high $k$ anticipated in RPT \cite{Crocce:2005xz} can be automatically reproduced if desired by using the exact initial conditions in phase-space.  

Next, let us comment on the relation between the extended BBGKY and the Helm\-holtz hierarchies. Combining the extended BBGKY equation for the linear response, (\ref{fIzeta}), with (\ref{responsecumulant}), (\ref{Wcum}), and (\ref{3vert}) one can recover the second Helmholtz equation, eq.~(\ref{responseFun}) above. From the extended BBGKY equation for the 2-pt function, (\ref{proba3}), using the same procedure, after a bit of algebra one can recover (\ref{2ptFun}).

This relationship between the first couple of equations of the (extended) BBGKY hierarchy and the HH should not be surprising. The (extended) BBGKY hierarchy is simply the hierarchy governing the evolution of $W^{(n)}$, while the HH governs the evolution of $\Gamma^{(n)}$. As $W$ and $\Gamma$ are related by a Legendre transformation, the two hierarchies must lead to identical dynamics.

Therefore, it should be possible to extract a closure relation for the extended BBGKY hierarchy starting from the truncated HH. Using (\ref{3vert}), one can start from the 3-vertex, $\Gamma^{(3)}$, to obtain the three-point correlation function $\langle f_af_bf_c\rangle_{\mathrm{c}}$ of the CDM, and the two-point linear response, $\delta\langle f_a f_b\rangle_{\mathrm{c}}/\delta \zeta_c$. Dropping the $\mathcal{O}(\varepsilon^3)$ terms which include diagrams with vertices $\Gamma_{\bar \chi\bar \chi \bar f}$ and $\Gamma_{\bar \chi\bar \chi \bar \chi}$, for the 3-pt function we obtain:
\newline
\begin{myfigure}
\includegraphics[scale=1.18]{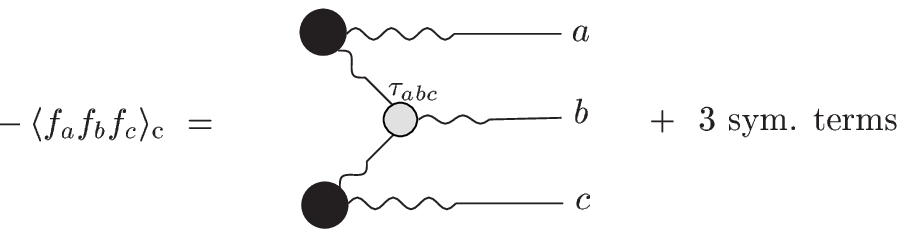}
\caption{\label{closure_3corr}}
\end{myfigure}
\noindent where we made explicit the value of $\tau$ for the vertices $\Gamma^{(3)}$. The 2-pt linear response functions is given by:
\vspace{0.3cm}
\begin{myfigure}
\includegraphics[scale=1.18]{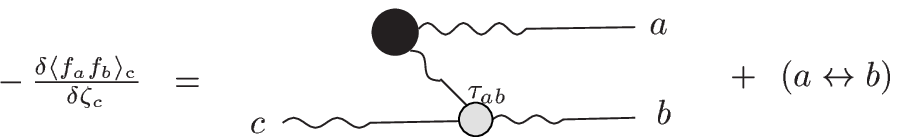}
\caption{\label{closure_2resp}}
\end{myfigure}
\noindent If one is to re-include the diagrams with vertices $\Gamma_{\bar \chi\bar \chi \bar f}$ and $\Gamma_{\bar \chi\bar \chi \bar \chi}$ in the above two equations, the equations become exact.

The interpretation of the closure relations above is straightforward. The closure relation for the 3-pt function, (\ref{closure_3corr}), tells us that a three point function is generated by the gravitational coupling (gray vertex) between two modes (which originate from two distinct pairs of modes, generated at the black blobs), the result of which is then fed into the response function, which propagates the effect to a third mode. Meanwhile, the closure relation for the 2-point linear response function, (\ref{closure_2resp}), tells us that under a non-random forcing, the 2-pt function varies because of the coupling of the linear response to a second mode (coming from a pair of modes generated at the black blob).

Therefore, the Helmholtz hierarchy collapses to the extended BBGKY hierarchy, (equations (\ref{proba2}), (\ref{proba3}) and (\ref{fIzeta})), with closure relations given by the three-point function:
\be\label{closure3pt}
\la f_af_bf_c \ra_\c&=&2 \RRl{a}{x}K_{xyz}G_{yb}G_{zc}\ + \ \hbox{3 sym. terms}\ +\ \mathcal{O}(\varepsilon^3)\nonumber\\
&=&2 \frac{\delta \bar  f_a}{\delta\zeta_x}K_{xyz}\la f_yf_b\ra_\c\la f_zf_c\ra_\c\ + \ \hbox{3 sym. terms}\ +\ \mathcal{O}(\varepsilon^3)\ ,
\ee
and the two-point linear response function:
\be\label{closure2ptresponse}
\frac{\delta\la f_af_b \ra_\c}{\delta\zeta_c}&=&2 \RRl{a}{x}K_{xyz}G_{yb}\RRl{z}{c}\  + \ (a\leftrightarrow b)\ +\ \mathcal{O}(\varepsilon^3)\nonumber\\
&=&2 \frac{\delta \bar f_a}{\delta\zeta_x}K_{xyz}\la f_yf_b\ra_\c\frac{\delta \bar f_z}{\delta\zeta_c}\  + \ (a\leftrightarrow b)\ +\ \mathcal{O}(\varepsilon^3)\ .
\ee
Note, however, that the truncation of $\Gamma$ at $\mathcal{O}(\varepsilon^3)$ still generates all possible correlation and response functions which enter the extended BBGKY hierarchy. Thus, the BBGKY hierarchy is not truncated (there is no $n$ above which $\langle f^n\rangle_{\mathrm{c}}$ is zero) under the closure relations above.

\section{Restoring the non-Gaussianities in $f_I$}\label{app:ngf}

So far we assumed that $f_I$ is a Gaussian random field. However, as we discussed in Section~\ref{sec:zeldovich_IC} and as we will see below keeping the non-Gaussian part of $f_I$ will turn out to be extremely important if one wants the expansion of the Helmholtz hierarchy for the CDM in $\varepsilon$ to be self-consistent.

Restoring the non-Gaussian part of $f_I$ is rather straightforward. The starting point again is the ensemble average of a functional, $F[f]$, which can be written as:
\be\label{FbegiNG}
\la F[f]\ra&=&\int \dD f_I \,F[f] p(f_I)=\int \dD f_I \,F[f] \int \dD y e^{i y f_I}\int \dD f_I' e^{-i y f_I'}p(f_I')\nonumber\\
&=&\int \dD f_I \,F[f] \int \dD y e^{i y f_I}\la e^{-i y f_I'}\ra\ ,
\ee
where $p(f_I)$ is the probability distribution from which $f_I$ is drawn. 
The term in the angular brackets above is simply $\exp( W_I[-i y])$, where $W_I$ is the Gibbs free energy for $f_I$ which can be written as:
\be
W_I[j]=\sum\limits_{n=1}^\infty \frac{j^n}{n!}\la f_I^n\ra_\c\ .
\ee

Performing the same manipulations on eq.~(\ref{FbegiNG}) as we did on eq.~(\ref{Fbegi}) in Section~\ref{sec:piVP}, we obtain
\be
\la F[f]\ra&=&\int \dD \phi F[f] e^{-S[\phi]}\ , \ \mathrm{where}\label{actionNG} \\
S[\phi]&\equiv&\chi_a \L{a}{b}f_b-\chi_a  K_{abc}f_bf_c-\chi_a \mu_a-\frac{1}{2}\chi_a\Delta_{ab}\chi_b-\label{SNG}\\\nonumber
&&-\frac{1}{3!}\chi_a\chi_b\chi_c\la f_{I,a}f_{I,b}f_{I,c}\ra_\c
-\frac{1}{4!}\chi_a\chi_b\chi_c\chi_d\la f_{I,a}f_{I,b}f_{I,c}f_{I,d}\ra_\c-\dots
\ ,
\ee
where the first line of the expression for the bare action $S$ corresponds identically to eq.~(\ref{S}). The bare action above has all the necessary properties that we required in Sections~\ref{sec:causalityI} and \ref{sec:causalityII} in order for the HH to lead to self-consistent and causal dynamics.

Note that now we have modified initial conditions for $\Gamma_\t$, sich that we have non-zero $\delta^n\Gamma_{I}/\delta \chi^n$ for all $n$. This modifies the power counting in $\varepsilon$. However, this effect is still perturbative and tractable. The reason for that is because in a diagram for a given $n$-point function the legs of $\delta^n\Gamma_{I}/\delta \chi^n$ must be attached to the propagator $\RR{a}{b}$, which in turn can correspond to either external legs or must attach to the bare vertex $K$ which is $\mathcal{O}(\varepsilon)$. If all $\RR{a}{b}$ correspond to external legs then to zero order that term is simply the $n$-point function in the ZA. Therefore, a given $n$-point function is given by the corresponding $n$-point function in the ZA plus higher order corrections. This is reasonable, as we explicitly chose the ZA as our zero-order solution.

Note that none of Equations (\ref{2pt_corr})-(\ref{SIGMA})
 except eq.~(\ref{pi})  are modified, as they explicitly omit the $\Gamma_{\bar\chi\bar\chi\bar\chi}$ term which arises from the non-Gaussian contributions to $f_I$. Truncating those equations at third order we find that only the self-energy receives a modification from non-Gaussianities in $f_I$ as follows:
\be\label{piNG}
\Pi_{xy}=4 K_{x12}G_{13}G_{24}K_{y34}+ 4 \la f_{I,a}f_{I,x}f_{I,y}\ra_\c \RR{x}{u}\RR{y}{v} K_{zuv}\RR{z}{b}\ .
\ee

Working to zero order in $\mathcal{\varepsilon}$ we again recover the ZA. At first order we obtain
\be
\RR{a}{b}^{(1)}=\RR{a}{x}^{(0)}\left(2K_{zxy}\bar f_y^{(0)}-L_{zx}^{(1)}\right)\RR{z}{b}^{(0)}\ ,\\
G^{(1)}_{ab}=2G_{ax}^{(0)}\left(2K_{zxy}\bar f_y^{(0)}-L_{zx}^{(1)}\right)\RR{z}{b}^{(0)} +2 \la f_{a}f_{x}f_{y}\ra_\c^{(0)} K_{zxy}\RR{z}{b}^{(0)}\ ,\label{GNG}
\ee
where the expression for $G_{ab}^{(1)}$ must be symmetrized. Here a  subscript in parenthesis denotes the order in the $\varepsilon$-expansion to which we work in. The first order quantity, $L^{(1)}$, is the term in $L$ which contains $\o\cdot\partial_{\o}$. The last term in the expression for $G^{(1)}_{ab}$ is due to the non-Gaussian contributions from $f_I$. Note that all zero-order quantities are evaluated in the ZA.

We can also write the closure relations for the BBGKY hierarchy including the non-Gaussian contributions from $f_I$. Both eq.~(\ref{closure_2resp}) and (\ref{closure2ptresponse}) remain unchanged, except they now have corrections of order $\mathcal{O}(\varepsilon^3)$. We recover eq.~(\ref{closure3pt}) but with the addition of the initial 3-point function which can be evaluated in the ZA:
\be\label{closure3ptNG}
\la f_af_bf_c \ra_\c&=&\RRl{a}{x}\RRl{b}{y}\RRl{c}{z}\la f_{I,x}f_{I,y}f_{I,z} \ra_\c+\\
&&+\left(2 \RRl{a}{x}K_{xyz}G_{yb}G_{zc}\ + \ \hbox{3 sym. terms}\right)\ +\ \mathcal{O}(\varepsilon^2)\nonumber\\
&=&\frac{\delta \bar  f_a}{\delta\zeta_x}\frac{\delta \bar  f_b}{\delta\zeta_y}\frac{\delta \bar  f_c}{\delta\zeta_z}\la f_{I,x}f_{I,y}f_{I,z} \ra_\c+\nonumber
\\
&&+\left(2 \frac{\delta \bar  f_a}{\delta\zeta_x}K_{xyz}\la f_yf_b\ra_\c\la f_zf_c\ra_\c\ + \ \hbox{3 sym. terms}\right)\ +\ \mathcal{O}(\varepsilon^2)\ .\nonumber
\ee
Similarly, one needs to modify eq.~(\ref{closure_3corr}).
Note that when evaluated to zero order in $\varepsilon$ the above equation reduces to the Zel'dovich 3-pt function.

\section{Particle interpretation}\label{PIinterp}

The solutions for the 2-pt function provided in the previous section for different orders in $\varepsilon$ are numerically hard to solve, because of the phase-space $n$-pt functions in the ZA.

Therefore, in this section we will rewrite our results in terms of effective particle trajectories. Those would allow one to run N-body simulations which will recover the different orders in $\varepsilon$. In order to proceed, note that one can write the exact VP equation as:
\be\label{f_Born_like}
f_x=\RRl{a}{z}^{(0)}\left[K_{zxy}f_xf_y-L_{zx}^{(1)}f_x\right]\ .
\ee
Using a Born-like approximation one can solve the above equation iteratively by starting with a solution which is close to the true solution for $f$. Choosing the initial $f$ for the Born-like approximation to be $f$ in the ZA, one can obtain a solution for $f$. That solution should have identical statistics as the ones obtained through the Helmholtz hierarchy, which uses the same expansion parameter as the Born-like approximation in the equation above.

Indeed at first order eq.~(\ref{f_Born_like}) reduces identically to the first order result in the HH, eq.~(\ref{GNG}). After some algebra at first order the above equation yields
\begin{flalign}\label{f1HH}
&f^{(1)}(\bm{x},\bm{v},\eta)=\int d^3q\int\limits_0^{\eta}d\eta'\frac{1}{D'}\frac{d\ln(a'\dot D')}{d\eta'}\times\\
\nonumber
&\times\left[\bigg(\partial_{{\bm x}'} \Phi^{(0)}(\bm{x}',\eta')+D'\bm{s}(\bm{q})\bigg)\cdot\partial_{\bm v}+(D-D')\bigg(\partial_{{\bm x}'} \Phi^{(0)}(\bm{x}',\eta')+D'\bm{s}(\bm{q})\bigg)\cdot\partial_{\bm x}\right]\\
\nonumber
&\ \ \ \ \ \d\left(\bm{x}-\bm{q}-D\bm{s}(\bm{q})\right)\d\left(\bm{v}-\bm{s}(\bm{q})\right)\ ,
\end{flalign}
where $\bm{x}'\equiv\bm{q}+D'\bm{s}(\bm{q})$, and primed quantities are evaluated at $\eta'$. And $\Phi$ is defined as 
$$\nabla^2\Phi\equiv\delta\ .$$
Note that to first order in $\delta$, we have $\partial_{\bm x}\Phi=-D\bm{s}$, and therefore $f^{(1)}$ above contains contributions which are only second order and higher in the density perturbations. Moreover, this ensures that the time integral above converges. 

If we chose to not restore the non-Gaussian contributions to $f_I$ in the first-order equation in the HH, eq.~(\ref{GNG}), then one can check that this is equivalent to using an inconsistent $\Phi$ which contains a zero order piece in the overdensity. This in turn would make the time-integral above divergent, and the HH would have broken down even at first order. Thus, keeping the non-Gaussian contributions to $f_I$ is crucial for the self-consistency of the HH.

We can extract an equation of motion for particles which are described by the first order one-particle distribution above. To do that let us write their trajectories as
\be
\bm{x}(\bm{q},\eta)=\bm{q}+D\bm{s}(\bm{q})+\bm{y}(\bm{q},\eta)\ .
\ee
The corresponding one-particle distribution function has the form:
\begin{flalign}
f(\bm{x},\bm{v},\eta)=\int d^3 q\, \d\left(\bm{x}-\bm{q}-D\bm{s}(\bm{q})-\bm{y}(\bm{q},\eta)\right)\d\left(\bm{v}-\bm{s}(\bm{q})-\frac{\dot{\bm{y}}(\bm{q},\eta)}{\dot D}\right).
\end{flalign}
Expanding the above expression to first order in $y$ (denoted by a superscript $(1_y)$) we obtain
\begin{flalign}
f^{(1_y)}(\bm{x},\bm{v},\eta)= - &\int d^3 q\, \left(\bm{y}(\bm{q},\eta)\cdot \partial_{\bm{x}}+\frac{\dot{\bm{y}}(\bm{q},\eta)}{\dot D}\cdot  \partial_{\bm{v}}\right)\\
&\quad\quad\quad\quad\quad\quad\quad\quad\quad \d\left(\bm{x}-\bm{q}-D\bm{s}(\bm{q})\right)\d\left(\bm{v}-\bm{s}(\bm{q})\right)\ .\nonumber
\end{flalign}
Comparing $f^{(1_y)}$ in the above equation with $f^{(1)}$ from eq.~(\ref{f1HH}) we can read off both $\bm{y}(\bm{q},\eta)$ and $\dot{\bm{y}}(\bm{q},\eta)$, which are consistent with one another. So, indeed the HH hierarchy at first order can be treated by performing an N-body simulation using particles with trajectories\footnote{Note that when we extracted the relevant particle trajectories, we assumed an expansion in the displacement, $y$, with respect to the particle trajectories in the ZA. Thus, eq.~(\ref{eomXZ1}) will in fact lead to a one particle distribution function which features additional resummations in $y$ with respect to the first order in HH.}:
\be\label{eomXZ1}
\bm{x}(\bm{q},\eta)=\bm{q}+D\bm{s}(\bm{q})-\int\limits_0^\eta d\eta' (D-D') 
\frac{1}{D'}\frac{d\ln(a'\dot D')}{d\eta'}\bigg(\partial_{{\bm x}'} \Phi^{(0)}(\bm{x}',\eta')+D'\bm{s}(\bm{q})\bigg)\ ,\nonumber\\
\ee
where $\bm{x}'=\bm{q}+D'\bm{s}(\bm{q})$.
This $\bm{x}(\bm{q},\eta)$ is a solution to the following equation of motion
\begin{flalign}\label{eomYraw}
&\ddot{\bm{x}}-\frac{\ddot D}{\dot D}\dot{\bm x}=-\frac{\dot D}{D}\frac{d\ln(a\dot D)}{d\eta}\bigg(\partial_{{\bm x}} \Phi^{(0)}(\bm{x},\eta)+D\bm{s}(\bm{q})\bigg)
\hbox{ , or equivalently:}
\\
\label{eomY}
&\ddot{\bm{x}}+\mathcal{H}\dot{\bm x}+\partial_{\bm x}\phi^{(0)}=-a^{-1}\frac{d(a\dot D)}{d\eta}\int\limits_0^\eta d\eta'\frac{1}{D'}\frac{d\ln(a'\dot D')}{d\eta'}\bigg(\partial_{{\bm x}'} \Phi^{(0)}(\bm{x}',\eta')+D'\bm{s}(\bm{q})\bigg)\ ,
\end{flalign}
where we re-expressed $\Phi$ with the Newtonian gravitational potential $\phi$ on the left hand side of eq.~(\ref{eomY}). One can compare eq.~(\ref{eomY}) with the true equation of motion given by 
\be\label{eomTrue}
\ddot{\bm{x}}+\mathcal{H}\dot{\bm x}+\partial_{\bm x}\phi=0\ .
\ee
Thus, the right hand side of eq.(\ref{eomY}), which is first order in $\varepsilon$ and second order and higher in the overdensity, corrects for the fact that the potential $\phi$ on the left hand side is evaluated in the Zel'dovich approximation.
 
The above equation of motion, eq.~(\ref{eomYraw}) or (\ref{eomY}), is not very convenient for implementing in an N-body simulation, both because of the non-standard velocity it uses (cf. eq.~(\ref{eomYraw})), and because generalizing it to higher orders is not straightforward. One can modify a bit the expansion scheme above to remove both of these problems. This is simply achieved by replacing $(D-D')$ in eq.~(\ref{eomXZ1}) by $\int^\eta_{\eta'}d\eta''1/a(\eta'')$ and dropping the logarithm:
\be\label{eomXZ1exact}
\bm{x}(\bm{q},\eta)&=&\bm{q}+D\bm{s}(\bm{q})-\\\nonumber
&&-\int\limits_0^\eta d\eta' \left(\int^\eta_{\eta'}d\eta''\frac{1}{a(\eta'')}\right) 
\frac{1}{D'}\frac{d(a'\dot D')}{d\eta'}\bigg(\partial_{{\bm x}'} \Phi^{(0)}(\bm{x}',\eta')+D'\bm{s}(\bm{q})\bigg)\ .
\ee
One can check that when evaluated with the full $\Phi$, and with the exact $\bm{x}'$ (i.e. not with $\bm{x}'$ evaluated in the ZA), the above equation is the exact solution to the true equation of motion, eq.~(\ref{eomTrue}). This solution is written in a form which contains explicitly the ZA, and is manifestly rather close to the $\varepsilon$-expansion\footnote{Indeed, starting from the VP equation and expanding in our variable $y$ one can show that the two are related by resumming certain higher order terms in $\varepsilon$.}.

Writing the equation of motion eq.~(\ref{eomTrue}) and its solution eq.~(\ref{eomXZ1exact}) to higher orders in $(\nabla\Phi+D\bm{s})$ is straightforward:
\be\label{eomNorder}
\ddot{\bm{x}}^{\leqslant n}+\mathcal{H}\dot{\bm x}^{\leqslant n}=-\partial_{\bm x}\phi^{\leqslant (n-1)}(\bm{x}^{\leqslant m})\ ,
\ee
where a subscript $(\leqslant n)$ implies that the quantity is a sum of all orders up to and including $n$. Above one has a choice of $m=n$ or $m=n-1$. If we are to stick as close as possible to the HH, then we should choose $m=n-1$. However, $m=n$ may offer faster convergence -- this is something which needs to be investigated further numerically.  The zero order is given by $\phi^{(0)}$ and $\bm{x}^{(0)}$ in the ZA.

Equation (\ref{eomNorder}) represents an iterative scheme for improving the Zel'dovich approximation. It is inspired by the $\varepsilon$ expansion, and if required can be modified to conform exactly to the $\varepsilon$ expansion in the same way in which we obtained the first order solution, eq.~(\ref{eomXZ1}). 
 
The above iterative scheme for solving the VP equation is well suited for implementing in N-body codes. This can be achieved in the way outlined in Section~\ref{sec:introd}.
 
\section{Summary}\label{sec:summary}

In this paper we derived a self-consistent hierarchy (the Helmholtz hierarchy) of partial differential equations, governing the evolution of the phase-space statistics of cold dark matter in the absence of baryons. The Helmholtz hierarchy (HH) has a physical ordering parameter, which interpolates between Zel'dovich dynamics and fully-fledged gravitational dynamics. It is schematically given by the fractional difference between the acceleration of test particles as given by a Zel'dovich-type approximation (i.e. the acceleration is assumed parallel to the velocity at an intermediate moment in time), and their corresponding true acceleration due to gravity. 

We showed that the HH preserves information about stream crossing and we argued that it automatically generates the decay expected at high $k$ of the \textit{density} response function \citep{Crocce:2005xz}. However, unlike RPT and analogous approaches, the resulting non-linear power spectrum is expanded around the ZA, possibly allowing us to construct better models for the mildly non-linear (and possibly non-linear) regime.

Under a sharp truncation of the HH all $n$-point correlation functions of CDM are generated, in contrast to the BBGKY hierarchy. Combining this result with the presence of a physical ordering parameter, we find that the HH ameliorates the closure problem of the BBGKY hierarchy.

We derived the HH for CDM using the functional renormalization group with a temporal cutoff \cite{Gasenzer:2007}, so that causality is built-in from the start. We proved that causality is indeed preserved to all orders --- a nontrivial result\footnote{The usual calculations in statistical mechanics assume that the system under study is in equilibrium. In such cases, the resulting equations of motion usually have either no dependence on a time parameter, or causality is not manifest.\label{foot:equil}}, which is important since structure formation is an inherently out-of-equilibrium process.

The HH has several advantages over performing a numerical N-body simulation and then taking spatial averages to reproduce the ensemble averages. First, the correlation and response functions are smooth (over the domain where they are nonzero), unlike the one-particle distribution function, or in the case of N-body simulations, the density field. Second, the effect of the short-scale modes on the long-scale modes (and vice versa) is readily taken into account by the HH. Third, the HH may offer the opportunity to develop better analytical or semi-analytical templates for fitting the nonlinear part of the CDM power spectrum obtained from numerical simulations.

The HH hierarchy is easy to solve numerically at zero order, since it corresponds identically to the ZA. However, going to higher orders requires a good numerical handle of the full phase-space $n$-point functions in the Zel'dovich approximation. To go around this problem, we proposed an iterative scheme which closely follows the $\varepsilon$-expansion. The scheme uses successive N-body simulations which improve upon the Zel'dovich approximation. However, we postpone such a numerical investigation to a future study.

\appendix
\section{Simplifying the contribution of $\Gamma^{(n+2)}$ to $\Gamma^{(n)}$}\label{app:zeroblob}

In the solutions to the NPRG hierarchy with a temporal cutoff, the time parameters in all 1PI diagrams must come earlier than the external legs of the corresponding 1PI diagram. This comes as a direct consequence of the delta function in (\ref{cutoff2}). Examples of this can be seen in eq.s~(\ref{hie2a},\ref{hie3}). 

Thus, the following equation holds for effective actions satisfying (\ref{assumed_for_now}), (\ref{n_vert_Gamma}) and (\ref{Rab_cause}):
\newline
\begin{myfigure}
\includegraphics[scale=1.2]{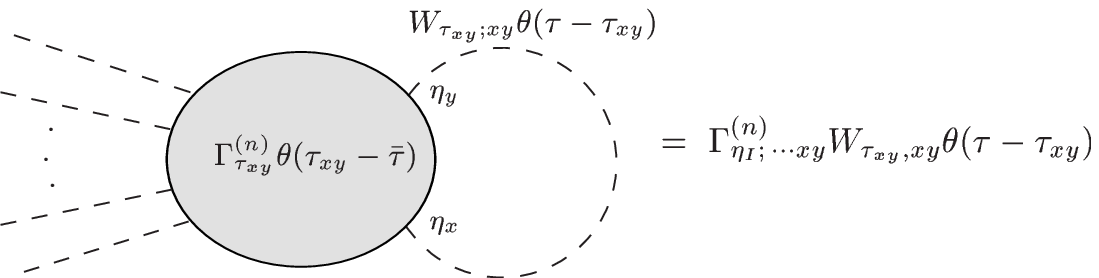}
\caption{\label{blobequation}}
\end{myfigure}
\noindent where $\bar\tau$ is the maximum of the time parameters running inside $\Gamma^{(n)}$; dashed lines represent either wiggly or straight lines. To derive the above equation we used the causality conditions (\ref{n_vert_Gamma}) combined with (\ref{wiggly_prop}), and the $\theta$-function restrictions written explicitly in the Feynman diagram above.
Up to a proportionality constant, the left hand side is the general form of the $\Gamma^{(n=m+2)}$ term in the equation for $\Gamma^{(m)}$. Therefore, using $\Gamma^{(n\geq4)}_{\eta_I}=0$ for the CDM, from (\ref{blobequation}) we find that the $\Gamma^{(n+2)}_\t$ term vanishes in the expression for $\Gamma^{(n)}_\t$ for $n>1$.

\acknowledgments The author would like to thank Matias Zaldarriaga for many helpful conversations.

\end{document}